\documentclass[aps,twocolumn,preprintnumbers,amsmath,amssymb,nofootinbib,superscriptaddress,notitlepage]{revtex4-2}

\usepackage[utf8]{inputenc}
\usepackage{braket}
\usepackage{bm}
\usepackage{slashed}
\usepackage{mathrsfs}
\usepackage{graphicx}
\usepackage{booktabs}
\usepackage{multirow}
\usepackage[colorlinks=true,linkcolor=blue,citecolor=blue,urlcolor=blue]{hyperref}
\usepackage{booktabs} 

\begin{document}

\title{Short-range dynamics of $X(3872)$ and $T_{cc}(3875)^+$ in a coupled-channel framework}

\author{Fang-Zheng Peng}


\affiliation{TUM School of Natural Sciences, Physics Department, Technical University of Munich, Garching 85748, Germany}

\author{Li-Sheng Geng}~\email{lisheng.geng@buaa.edu.cn}
\affiliation{Sino-French Carbon Neutrality Research Center, \'{E}cole Centrale de P\'{e}kin/School of General Engineering, Beihang University, Beijing 100191, China}
\affiliation{School of Physics, Beihang University, Beijing 102206, China}%
\affiliation{Peng Huanwu Collaborative Center for Research and Education, Beihang University, Beijing 100191, China}
\affiliation{Southern Center for Nuclear-Science Theory (SCNT), Institute of Modern Physics, Chinese Academy of Sciences, Huizhou 516000, China}%

\author{Ju-Jun Xie}~\email{xiejujun@impcas.ac.cn}
\affiliation{Southern Center for Nuclear-Science Theory (SCNT), Institute of Modern Physics, Chinese Academy of Sciences, Huizhou 516000, China}%
\affiliation{Heavy Ion Science and Technology Key Laboratory, Institute of Modern Physics, Chinese Academy of Sciences, Lanzhou 730000, China}%
\affiliation{School of Nuclear Sciences and Technology, University of Chinese Academy of Sciences, Beijing 101408, China}

\date{\today}

\begin{abstract}

We study the short-range dynamics in the coupled-channel description of the $S$-wave $D \bar{D}^*$ and $DD^*$ systems for $X(3872)$ and $T_{cc}(3875)^+$, respectively. Contact interactions are related through scalar- and vector-meson exchange saturation, while pion exchange is included explicitly with its local term subtracted. The coupled Lippmann-Schwinger equations incorporate the self-energy of the unstable $D^*$ mesons, thereby accounting for the three-body $D \bar{D} \pi$ ($D D\pi$) dynamics. Fixing the contact interaction from the $T_{cc}(3875)^+$ pole, we trace the predicted $X(3872)$ pole as the local magnetic dipole term from vector-meson exchange is varied. Retaining this term in full produces the shallowest poles, which remain approximately $5$--$12~\mathrm{MeV}$ below the observed $X(3872)$ mass. It is found that the pion exchange has a moderate effect on the binding energy and a stronger effect on the isospin violation for the $X(3872)$ state. The residual tension between the pure-molecular bound descriptions of $X(3872)$ and $T_{cc}(3875)^+$ is analyzed with the Feshbach projection. This analysis indicates a $DD^*$ interaction weaker than that implied by the purely molecular fits, and shows that the $T_{cc}(3875)^+$ state instead arises from the mixing of the $DD^{*}$ channel with a compact $cc\bar{u}\bar{d}$ state. We further examine the spatial structure using Born--Oppenheimer effective field theory. For separation radii of $0.5$--$1.5~\mathrm{fm}$, the combined charmonium and short-distance tetraquark weight in $X(3872)$ is $3.0$--$9.7\%$, while the short-distance weight in $T_{cc}(3875)^+$ is $1.3$--$14.4\%$. These results favor retaining the local vector-exchange contribution within the saturation framework and motivate a joint treatment of molecular and compact dynamics for the $T_{cc}(3875)^+$.

\end{abstract}

\maketitle

\section{Introduction}

Exotic hadrons beyond the conventional quark-antiquark and three-quark classifications provide a testing ground for nonperturbative quantum
chromodynamics~\cite{Brambilla:2010cs,Esposito:2016noz,Ali:2017jda,Liu:2024uxn}. Some exotic hadron candidates lie near two-hadron thresholds, where compact multiquark configurations and hadronic molecules may both contribute to such observed states~\cite{Chen:2016spr,Guo:2017jvc,Liu:2019zoy,Chen:2022asf,Meng:2022ozq}. Testing a molecular interpretation therefore requires controlling both the interaction between the components and the short-distance assumptions that enter it.

The one-boson-exchange (OBE) model provides a useful phenomenological description of hadron-hadron interactions~\cite{Machleidt:1987hj,Ding:2008gr,Liu:2019stu}. Derivative meson couplings can generate momentum-independent terms whose Fourier transforms are proportional to $\delta^{(3)}(\bm r)$; once regulated, these terms act over distances of the order of the inverse cutoff and can substantially affect near-threshold poles~\cite{CalleCordon:2009pit,PavonValderrama:2019nbk,Liu:2019zvb}. Molecular calculations have both retained and subtracted such local terms~\cite{Thomas:2008ja,Liu:2009qhy,Meng:2017fwb,Wang:2020dya,Chen:2021tip,Ling:2021asz,Wang:2022mxy,Wang:2024ukc}. Their treatment hinges on the separation between resolved meson exchange and unresolved short-distance physics, a distinction that becomes particularly relevant when the pion exchange is compared with the exchange of heavier mesons whose masses are comparable to the regulator scale. Heavy-meson chiral effective field theory (HMChEFT) implements this separation through contact low-energy constants (LECs) accompanied by explicit pion exchanges~\cite{PavonValderrama:2019ixb,Meng:2022ozq}. The effective theory alone does not relate the LECs of different molecular
systems. Meson-exchange saturation provides an additional model for them, matching the effects of the heavier exchanged mesons onto contact interactions~\cite{Wang:2022jop,Zou:2025sae,Yalikun:2025ssz,Wang:2025ioe}. A related prescription combines light-meson saturation with renormalization-group scaling, bringing contributions associated with different exchange masses to a common scale~\cite{Peng:2020xrf,Peng:2021hkr,Yang:2022ezl,Yan:2023ttx}. Within this prescription, the treatment of the local exchange terms can be tested through the correlations they induce between different hadronic systems.

The $X(3872)$ and $T_{cc}(3875)^+$ states are particularly well suited to such a comparison. The $X(3872)$ (also denoted as $\chi_{c1}(3872)$~\cite{ParticleDataGroup:2026mpi}) was discovered by the Belle Collaboration and has since been observed in several production and decay processes~\cite{Belle:2003nnu,D0:2004zmu,Belle:2006olv,BaBar:2010wfc,BESIII:2013fnz,ATLAS:2016kwu,BESIII:2019esk,LHCb:2020xds,CMS:2021znk,BESIII:2022bse}. It has quantum-numbers $J^{PC}=1^{++}$~\cite{LHCb:2013kgk} and lies close to the $D^0\bar D^{*0}$ mass threshold, and its decays into $J/\psi\rho^0$ and $J/\psi\omega$ provide a sensitive probe of isospin breaking~\cite{Gamermann:2009fv,Gamermann:2009uq,Hidalgo-Duque:2012rqv,Li:2012cs,Meng:2021kmi,Wang:2022vjm,Zhang:2024fxy,Ji:2025hjw,Dong:2026qbr}. The doubly charmed $T_{cc}(3875)^+$ was observed by the LHCb Collaboration with quantum-numbers consistent with $J^P=1^+$, and its pole in a unitarized line-shape analysis lies about $360~\mathrm{keV}$ below the $D^0 D^{*+}$ mass threshold~\cite{LHCb:2021vvq,LHCb:2021auc}. These threshold proximities motivate molecular descriptions of both the $X(3872)$ and $T_{cc}(3875)^+$ states~\cite{Braaten:2004fk,Lee:2009hy,Nieves:2011zz,Nieves:2012tt,Guo:2013zbw,Guo:2014hqa,Albaladejo:2015dsa,Kalashnikova:2018vkv,Contessi:2020jqa,Wang:2022xga,Esposito:2025hlp,Feijoo:2021ppq,Albaladejo:2021vln,Du:2021zzh,Yan:2021wdl}. A simultaneous molecular description is nevertheless nontrivial. Contact interactions saturated by $\sigma$, $\rho$, and $\omega$ exchanges, with the local vector-exchange $\delta^3(\bm r)$ terms removed, were found to produce tension between the two states in a single-channel analysis~\cite{Peng:2023lfw}. A phenomenological OBE study likewise found that the role of these terms depends on the system and on the chosen cutoff~\cite{Qiu:2023uno}. Moreover, since meson-antimeson and meson-meson LECs are independent in the general effective theory, correlations inferred from saturation should be tested as model assumptions rather than imposed as symmetry identities.

In this work, we examine the local term generated by the magnetic coupling of the exchanged vector mesons in a coupled-channel calculation. We include the neutral and charged channels of both systems, the energy-dependent $D^*$ self-energies, and the subtracted leading-order one-pion-exchange (OPE) potential. The real part of the $T_{cc}(3875)^+$ pole fixes the overall contact strength, after which the common saturation prescription determines the interaction in the $X(3872)$ sector. We study the resulting pole trajectories and the residue-based isospin-violating coupling ratio. To interpret the residual tension, we also consider the Feshbach reduction and the independent analysis of the short-distance wave-function weights in Born-Oppenheimer effective field theory (BOEFT). The Feshbach reduction shows that the need of a compact component for $T_{cc}(3875)^+$ is controlled by the short-distance dynamics, while the BOEFT wave functions quantify this component with the adopted separation scale.

The paper is organized as follows. Sec.~\ref{sec2: molecular channels and interactions} defines the coupled channels and the corresponding interactions, and Sec.~\ref{sec3: LS equation and sigularities} presents the scattering equation, its analytic continuation, and the isospin-violating coupling ratio. The numerical results, the Feshbach analysis, and the BOEFT wave functions are presented in Sec.~\ref{sec4: numerical results}, and we summarize and discuss our results in Sec.~\ref{sec5: summary and discussions}.

\section{Coupled molecular channels and interactions} \label{sec2: molecular channels and interactions}

\subsection{Physical channels for $X(3872)$ and $T_{cc}(3875)^+$}

Using the isospin convention, for the positive-charge-conjugation state $X(3872)$ with $J^{PC}=1^{++}$, the isoscalar and isovector $D \bar{D}^*$ channels are given by~\footnote{Here, the isospin conventions for quarks and antiquarks are
\begin{eqnarray}
    \ket{u}=\ket{\frac{1}{2},+\frac{1}{2}} \,, && \ket{d}=\ket{\frac{1}{2},-\frac{1}{2}}; \nonumber \\
   \ket{\bar{d}}=-\ket{\frac{1}{2},+\frac{1}{2}} \,, && \ket{\bar{u}}=\ket{\frac{1}{2},-\frac{1}{2}}. \nonumber
\end{eqnarray}}
\begin{eqnarray}
\begin{aligned}
&\ket{D\bar D^{*},I=0}
=\frac{1}{\sqrt{2}}(\bar u u+\bar d d)
  \otimes\frac{1}{\sqrt{2}}
  \ket{D\bar D^*-D^*\bar D} \\
&=\frac{1}{\sqrt{2}}\left(
  \frac{1}{\sqrt{2}} \ket{D^0\bar D^{*0}-D^{*0}\bar D^0} +
 \frac{1}{\sqrt{2}} \ket{D^+D^{*-}-D^{*+}D^-}
  \right), \\[3pt]
&\ket{D\bar D^{*},I=1}
=\frac{1}{\sqrt{2}}(\bar u u-\bar d d)
  \otimes\frac{1}{\sqrt{2}}
  \ket{D\bar D^*-D^*\bar D} \\
&=\frac{1}{\sqrt{2}}\left(
  \frac{1}{\sqrt{2}} \ket{D^0\bar D^{*0}-D^{*0}\bar D^0} -
 \frac{1}{\sqrt{2}} \ket{D^+D^{*-}-D^{*+}D^-}
  \right),
\label{eq:X-channels-isospin-01}
\end{aligned}
\end{eqnarray}
where we use the charge-conjugation convention $C\ket{D\bar{D}^*\pm D^*\bar{D}}=\mp\ket{D\bar{D}^*\pm D^*\bar{D}}$ as in Refs.~\cite{Hidalgo-Duque:2012rqv,Zhang:2024fxy}. 

For the $T_{cc}(3875)^+$ state, we consider the $DD^*$ system with $J^P=1^+$. The isoscalar and isovector combinations are~\cite{Qiu:2023uno,Ge:2026moy}
\begin{eqnarray}
\begin{aligned}
&\ket{D^{*}D, I=0}=\frac{1}{\sqrt{2}} \left(\bar{u}\bar{d}-\bar{d}\bar{u} \right)\otimes \frac{1}{\sqrt{2}}\ket{DD^*+D^*D}
\\
&=\frac{1}{\sqrt{2}}\left(\frac{1}{\sqrt{2}}\ket{D^0 D^{*+}-D^{*+} D^0} - \frac{1}{\sqrt{2}}\ket{D^+ D^{*0}-D^{*0} D^+} \right),
\\[3pt]
&\ket{D^{*}D, I=1}=
\frac{-1}{\sqrt{2}} \left(\bar{u}\bar{d}+\bar{d}\bar{u} \right)\otimes \frac{1}{\sqrt{2}}(\ket{DD^*+D^*D})
\\
&= \frac{-1}{\sqrt{2}}\left(\frac{1}{\sqrt{2}}\ket{D^0 D^{*+}+D^{*+} D^0} + \frac{1}{\sqrt{2}}\ket{D^+ D^{*0}+D^{*0} D^+} \right).
\label{eq:T channels of isospin 0 1}
\end{aligned}
\end{eqnarray}

The neutral and charged $C$-even channels used for the $X(3872)$ state are
\begin{eqnarray}
\begin{pmatrix} \,\,
\ket{N_X} \,\,\\
\,\,\ket{C_X}
\,\, \end{pmatrix} =\frac{1}{\sqrt{2}} \begin{pmatrix}\,\,
 \ket{D^{0}\bar{D}^{*0}}- \ket{D^{*0}\bar{D}^{0}} \\
 \,\,
\ket{D^{+}D^{*-}}-\ket{D^{*+}D^{-}} \,\,
\end{pmatrix} \, .
\label{eq:X physical channels}
\end{eqnarray}
While for $T_{cc}(3875)^+$, the two charge channels are
\begin{eqnarray}
\begin{pmatrix} \,\,
\ket{N_T} \,\,\\
\,\,\ket{C_T}
\,\, \end{pmatrix} = \begin{pmatrix}\,\,
 \ket{D^{0}D^{*+}} \\
 \,\,
\ket{D^{+}D^{*0}} \,\,
\end{pmatrix} ,
\label{eq:T physical channels}
\end{eqnarray}
with $\ket{N_T}$ and $\ket{C_T}$ the channels containing a neutral $D^0$ and a charged $D^+$ meson, respectively.

The above physical charge basis for $X(3872)$ is related to the isospin channels by
\begin{equation}
\begin{aligned}
\ket{N_X}&=
\frac{1}{\sqrt{2}} (\ket{D\bar{D}^{*} , I=0}+\ket{D\bar{D}^{*} , I=1}),
\\
\ket{C_X}&=
\frac{1}{\sqrt{2}} (\ket{D\bar{D}^{*} , I=0}-\ket{D\bar{D}^{*} , I=1})   \, .
\end{aligned}
\label{eq:X physical and isospin channels}
\end{equation}

And, for the $T_{cc}(3875)^+$ state, the corresponding relations are~\footnote{Next, throughout the paper, we use $[\bar{D}D^*]$ and $[DD^*]$ as shorthand for the coupled $X(3872)$ and $T_{cc}(3875)^+$ physical channels in Eqs.~\eqref{eq:X physical channels} and \eqref{eq:T physical channels}, respectively.}
\begin{equation}
\begin{aligned}
\ket{N_T}&=
-\frac{1}{\sqrt{2}} (\ket{DD^{*} , I=1}-\ket{DD^{*} , I=0}),
\\
\ket{C_T}&=
-\frac{1}{\sqrt{2}} (\ket{DD^{*} , I=1}+\ket{DD^{*} , I=0})  \, .
\end{aligned}
\label{eq:T physical and isospin channels}
\end{equation}

In heavy-meson chiral effective field theory (HMChEFT), the leading-order (LO) interaction $V$ between a charmed meson-(anti)meson pair can be expressed, in momentum space, as the sum of a contact potential and a OPE potential~\cite{PavonValderrama:2019ixb,PavonValderrama:2019nbk,Peng:2020hql,Peng:2023lfw,Meng:2022ozq}. For the physical heavy-meson pair channels of $X(3872)$ and $T_{cc}(3875)^+$ in Eqs.~\eqref{eq:X physical channels} and \eqref{eq:T physical channels}, we write the momentum-space interaction kernel as~\cite{Du:2021zzh,Zhang:2024fxy,Ji:2025hjw}

\begin{equation}
V(\bm p',\bm p) = V_{\rm Con} + V_{\rm OPE}(\bm p',\bm p),
\label{eq:V_contact_OPE}
\end{equation}
where $V_{\rm Con}$ parametrizes unresolved short-distance dynamics and $V_{\rm OPE}$ describes explicit one pion exchange. The separation between these terms requires a subtraction prescription, which we specify below.

\subsection{Long-range interaction from one-pion exchange}

The derivative $D^*D\pi$ and $\bar{D}^*\bar{D}\pi$ couplings follow from the LO heavy-meson chiral Lagrangian~\cite{Ding:2008gr,Liu:2019stu,Wang:2022jop},
\begin{eqnarray}
\begin{aligned}
    \mathcal{L}_{H\bar{H}\pi}^0=i g_{\pi} \left( \mathrm{Tr} [H^{Q}_b\slashed{A}_{ba}\gamma_5\bar{H}_a^{Q}]  
    + \mathrm{Tr} [\bar{H}^{\bar{Q}}_b\slashed{A}_{ba}\gamma_5 H_a^{\bar{Q}}] \right),
\end{aligned}
\end{eqnarray}
with $g_{\pi}=0.57$ determined from the  $D^*\rightarrow D\pi$ decay. The superfields $H^Q$ and $H^{\bar{Q}}$ annihilate mesons containing a heavy quark and a heavy antiquark, respectively:
\begin{eqnarray}
\begin{aligned}
H^{Q}&=& \frac{1+\slashed{\nu}}{2}\left(P_{\mu}^{\ast Q}\gamma^{\mu}- P^Q\gamma_5\right) \, , \,
\bar{H}^Q=\gamma^0 H^{Q\dagger}\gamma^0 ,  \\
H^{\bar{Q}}&=& \left(P_{\mu}^{\ast\bar{Q}}\gamma^{\mu}- P^{\bar{Q}}\gamma_5\right)\frac{1-\slashed{\nu}}{2} \, , \,
\bar{H}^{\bar{Q}}=\gamma^0 H^{{\bar{Q}}\dagger}\gamma^0 .
\end{aligned}
\end{eqnarray}

In the two-component notation, we have
\begin{eqnarray}
    P^{(*)Q}=(D^{(*)0}, D^{(*)+}) , ~~
    P^{(*)\bar{Q}}=(\bar{D}^{(*)0}, \bar{D}^{(*)-}).
\end{eqnarray}
The axial current $A_{\mu}=\frac{1}{2}(\xi^{\dagger}\partial_{\mu}\xi-\xi\partial_{\mu}\xi^{\dagger})$ can be expanded as $A_{\mu}=\frac{i}{2f_{\pi}}\partial_{\mu}\phi+\mathcal{O}(\phi^3)$, where $\xi=\mathrm{e}^{i\phi/(2f_\pi)}$, $f_{\pi}=92.1 \,\rm{MeV}$ is the pion decay constant~\cite{Hidalgo-Duque:2012rqv,Du:2021zzh}, and $\phi$ denotes the matrix of pseudoscalar meson fields,
\begin{eqnarray}
  \phi=
    \begin{pmatrix}
    \pi^0   &&\sqrt{2}\pi^+ \\
      \sqrt{2}\pi^-  &&-\pi^0
    \end{pmatrix} .
\end{eqnarray}

\begin{figure}[htbp]
    \centering
\includegraphics[width=\linewidth]{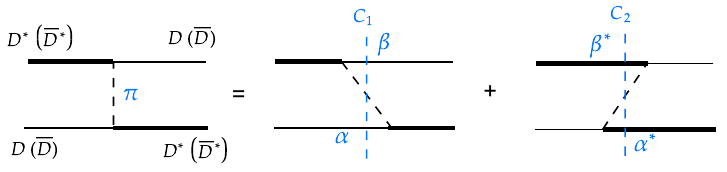}
    \caption{Time-ordered one-pion-exchange diagrams. The labels $\alpha,\beta$ ($\alpha^*,\beta^*$) denote pseudoscalar (vector) charmed mesons or antimesons. The dashed lines $C_1$ and $C_2$ mark the $\alpha\beta\pi$ and $\alpha^*\beta^*\pi$ intermediate states. Only $C_1$ can become an on-shell cut near the thresholds considered here.}
    \label{fig:OPE potential}
\end{figure}

Therefore, the generic structure of the OPE potential corresponding to Fig.~\ref{fig:OPE potential} can be expressed in momentum space as
\begin{equation}
V_{\rm OPE}(q) \propto -\frac{g_{\pi}^2}{4f_{\pi}^2} \frac{ (\epsilon^{\prime\dagger}\cdot q) (\epsilon\cdot q)} {q^2-m_\pi^2}, \label{eq:OPE_generic 1}
\end{equation}
where $q$ is the exchanged pion four-momentum and $\epsilon$ ($\epsilon'$) is the polarization vector of the initial (final) vector heavy meson. In the heavy-quark limit with four-velocity $\nu_{\mu}=(1,\bm 0)$, the transversality condition $\epsilon\cdot p=m_H\,\epsilon\cdot\nu=0$ for a massive vector meson of mass $m_H$ implies $\epsilon^0=0$. Therefore, by taking the energy transfer to be the $m_{D^*}-m_D$ mass difference, the OPE potential operator in Eq.~\eqref{eq:OPE_generic 1} can be simplified to a static expression
\begin{equation}
V_{\rm OPE}(\bm q) =\zeta \eta \, \bm{\tau}_1\cdot\bm{\tau}_2\frac{g_{\pi}^2}{4f_{\pi}^2} \frac{ (\bm\epsilon^{\prime\dagger}\cdot\bm q)
(\bm\epsilon\cdot\bm q)} {\bm q^2+\mu_\pi^2},
\label{eq:OPE_generic 2}
\end{equation}
with $\mu_{\pi}^2=m_{\pi}^2-\Delta_D^2$, where $\Delta_D$ denotes the mass splitting between $D$ and $D^*$. The factor $\eta$ accounts for the exchange symmetry of the meson-meson system or the charge-conjugation parity of the meson-antimeson system, and $\zeta$ is the $G$-parity factor. Their channel-dependent values are collected in Table~\ref{Table:lambda values}. The operator $\bm{\tau}_1\cdot\bm{\tau}_2$ denotes the scalar product of the isospin Pauli matrices of the two heavy mesons.

\begin{table}[htbp]
\renewcommand{\arraystretch}{1.3}
    \centering
   \caption{Channel factors entering the pion- and vector-meson-exchange potentials. The spin-parity assignment is $J^{PC}$ for $\bar{D}D^*$ and $J^P$ for $DD^*$.}
    \begin{tabular}{ccccccc}
        \toprule [1pt]
         &\,\, $J^{PC}$ & \,\,$I$ & \,\, $\zeta$ & $ \,\, \eta$ & $\,\, \tau_{\rm fac}$ & \,\, $S_{\rm fac}$\\
        \midrule [1pt]
        \multirow{2}{*}{$\bar{D}D^*$} &\,\, $1^{++}$ & \,\, $0$ & \,\, $1$  & \,\, $1$  & $-3$  & $1$
        \\
       & \,\, $1^{++}$ & \,\, $1$  & \,\, $1$ & \,\, $1$  & $1$ & $1$  \\
        \hline
       \multirow{2}{*}{$DD^*$}& \,\, $1^{+}$ & \,\, $0$  & \,\, $-1$ & \,\, $-1$  & $-3$ & $1$
       \\
      & \,\,  $1^{+}$ & \,\,  $1$  & \,\, $-1$ &\,\, $1$ & $1$ & $1$
      \\
        \bottomrule [1pt]
    \end{tabular}
     \label{Table:lambda values}
\end{table}

The separation between the local short- and long-distance components that emerges in Eq.~\eqref{eq:OPE_generic 2} is particularly relevant for near-threshold heavy-meson systems, for which the pion range is parametrically larger than the characteristic range associated with the ultraviolet regulator. To see this, note that the operator in the numerator of Eq.~\eqref{eq:OPE_generic 2} can be decomposed as~\cite{Valderrama:2012jv,Peng:2020xrf,Peng:2021hkr,Wang:2022jop}
\begin{eqnarray}
(\bm\epsilon^{\prime\dagger}\cdot\bm q)(\bm\epsilon\cdot\bm q)
=
\frac{1}{3}(\bm\epsilon^{\prime\dagger}\cdot\bm \epsilon)\bm{q}^2+\bm{q}^2S_{12},
\end{eqnarray}
where the tensor operator $S_{12}=(\bm\epsilon^{\prime\dagger}\cdot\hat{\bm q})(\bm\epsilon\cdot\hat{\bm q})-\frac{1}{3}(\bm\epsilon^{\prime\dagger}\cdot\bm \epsilon)$ vanishes for the $S$-wave molecular channels of $X(3872)$ and $T_{cc}(3875)^+$ considered throughout this paper. Then, the resulting central OPE potential becomes
\begin{equation}
V_{\rm OPE}(\bm q)
=\zeta \, \eta \, \tau_{\rm fac}\frac{g_{\pi}^2}{4f_{\pi}^2}\left(\frac{1}{3}-\frac{1}{3}\frac{\mu_{\pi}^2}{\bm{q}^2+\mu_{\pi}^2} \right) \, ,
\label{eq:OPE_generic delta}
\end{equation}
where the values of the isospin factor $\tau_{\rm fac}=\braket{\bm \tau_1\cdot \bm \tau_2}$ for the different quantum numbers of the molecular $H_1H_2$ system are listed in Table~\ref{Table:lambda values}. The constant term corresponds, after Fourier transformation, to a local contribution proportional to $\delta^{(3)}(\bm r)$, whereas the second term gives the finite-range Yukawa interaction. The local constant in Eq.~\eqref{eq:OPE_generic delta} originates from the use of a point-like $D^{(*)}D^{(*)}\pi$ vertex. Since the corresponding effective interaction is extrapolated to arbitrarily large momenta, the derivative pion coupling produces a momentum-independent component at large $|\bm q|$. In coordinate space this component is localized at zero separation. After regularization with an ultraviolet scale $\Lambda$, this delta function is smeared over a characteristic distance of order $r_{\pi}^{UV}\sim\frac{1}{\Lambda}$. For a typical hadronic cutoff satisfying $ \Lambda\gg m_\pi$, there is a clear hierarchy $r_{\pi}^{UV}\ll\frac{1}{m_\pi}$ between the regulator-localized contribution and the physical pion-exchange range. We therefore regard the local part generated by the point-like pion coupling as a short-distance artifact of the representation of OPE, rather than as part of the resolved long-range pion force.

We consequently define the subtracted OPE potential by directly removing this local contribution,
\begin{equation}
V_{\rm OPE}^{\rm sub}
=
V_{\rm OPE}^{\rm full}
-
V_{\rm OPE}^{\delta}.
\label{eq:OPE_subtracted}
\end{equation}
Here, $V_{\rm OPE}^{\delta}$ denotes the local $\delta^{(3)}(\bm r)$ component of the OPE. Thus, $V_{\rm OPE}^{\rm sub}$ contains the finite-range part of the pion exchange relevant for the long-distance dynamics. Our prescription amounts to a direct subtraction of the local component in momentum space. More elaborate semilocal regularization prescriptions can provide a smoother separation between the short- and long-distance components while preserving the long-range pion-exchange structure~\cite{Reinert:2017usi}.

Using the isospin transformations in Eqs.~\eqref{eq:X physical and isospin channels} and \eqref{eq:T physical and isospin channels}, the local $\delta^3(\bm r)$ contributions from the OPE potential in the ordered bases $(\ket{N_X},\ket{C_X})$ and $(\ket{N_T},\ket{C_T})$ are
\begin{eqnarray}
V_{\mathrm{OPE},\,X}^{\delta}&=&\frac{g_{\pi}^2}{12f_{\pi}^2}
    \begin{pmatrix}
       \, -1 & \,-2 \, \\
        \, -2 & \,-1 \,
    \end{pmatrix}  \, ,
    \\
V_{\mathrm{OPE},\,T}^{\delta}&=&\frac{g_{\pi}^2}{12f_{\pi}^2}
    \begin{pmatrix}
       \, -2 & \,1 \,\\
        \, 1 & \,-2 \,
    \end{pmatrix} \, .
\end{eqnarray}
For the Lippmann-Schwinger equation, we represent the full OPE potential $V_{\rm OPE}^{\rm full}$ in time-ordered perturbation theory (TOPT), rather than through the expression in Eq.~\eqref{eq:OPE_generic 2}, as shown by the two time-ordered diagrams in Fig.~\ref{fig:OPE potential}. Near the two-body thresholds, only the forward diagram (with cut $C_1$ in Fig.~\ref{fig:OPE potential}) can go on shell and thereby develops the $DD\pi$ ($\bar{D}\bar{D}\pi$) cut. The denominator $D_{\alpha\beta\pi}^{\rm full}$ of the full pion propagator can be decomposed into the forward ($D_{\alpha\beta\pi}$) and backward ($D_{\alpha^*\beta^*\pi}$) parts as~\cite{Du:2019pij,Qiu:2023uno,Zhang:2024fxy}
\begin{eqnarray}
  \frac{1}{D_{\alpha\beta\pi}^{\rm full}}=\frac{1}{2E_{\pi}(\bm q)}\left( \frac{1}{D_{\alpha\beta\pi}}+\frac{1}{D_{\alpha^*\beta^*\pi}} \right),
  \label{eq:full propagator of forward and backward types}
\end{eqnarray}
where
\begin{eqnarray}
    D_{\alpha\beta\pi} &=& E-E_{\alpha}(\bm p')-E_{\beta}(\bm p)-E_{\pi}(\bm q)+i\epsilon \, ,  \\
    D_{\alpha^*\beta^*\pi} &=& E-E_{\alpha^*}(\bm p)-E_{\beta^*}(\bm p')-E_{\pi}(\bm q)+i\epsilon \, .
\end{eqnarray}
Here $E$ is the total center-of-mass energy of the hadronic molecular system $H_1H_2$, $E_\pi(\bm q)=\sqrt{\bm q^2+m_\pi^2}$, and $\bm p$ ($\bm p'$) is the initial (final) relative momentum. The momentum transfer is $\bm q=\bm p-\bm p'$, and $z=\cos\theta$ is the cosine of the angle between $\bm p$ and $\bm p'$.

The full OPE potentials for the pure $S$-wave molecular interpretations of $X(3872)$ and $T_{cc}(3875)^+$ in the physical channels of Eqs.~\eqref{eq:X physical channels} and \eqref{eq:T physical channels} are then given by
\begin{eqnarray}
\begin{aligned}
V_{\mathrm{OPE},\,{X}}^{\rm full}(E;\bm p', \bm p) =
\frac{g_{\pi}^2}{12 f_{\pi}^2}
\begin{pmatrix}
     V_{D^0\bar{D}^0\pi^0} & \, 2V_{D^0D^-\pi^+}  \\
     2V_{D^+\bar{D}^0\pi^-} & \, V_{D^+D^-\pi^0} 
\end{pmatrix} ,
\\[3pt]
V_{\mathrm{OPE},\,{T}}^{\rm full} (E;\bm p', \bm p)=
\frac{g_{\pi}^2}{12 f_{\pi}^2}
\begin{pmatrix}
    2V_{D^0D^0\pi^+} & \, -V_{D^0D^+\pi^0}  \\
     -V_{D^+D^0\pi^0} & \, 2V_{D^+D^+\pi^-} 
\end{pmatrix},
\end{aligned}
\end{eqnarray}
where
\begin{eqnarray}
    V_{\alpha\beta\pi}=\frac{1}{2}\int_{-1}^1 \frac{\bm q^2}{D_{\alpha\beta\pi}^{\rm full}(E;\bm p', \bm p, z)} dz \, .
    \label{eq: OPE potential partial waves}
\end{eqnarray}

Therefore, the subtracted OPE potentials for $X(3872)$ and $T_{cc}(3875)^+$ read
\begin{eqnarray}
\begin{aligned}
 V_{\mathrm{OPE}, \, X}^{\rm sub}
=
V_{\mathrm{OPE}, \, X}^{\rm full}
-
V_{\mathrm{OPE}, \, X}^{\delta} \,,
\\[3pt]
 V_{\mathrm{OPE}, \, T}^{\rm sub}
=
V_{\mathrm{OPE}, \, T}^{\rm full}
-
V_{\mathrm{OPE}, \, T}^{\delta} \,\,.
\end{aligned}
\label{eq: sub-OPE for X and T}
\end{eqnarray}

\subsection{Short-range contact interactions saturated by meson exchanges}

In the physical basis of Eq.~\eqref{eq:X physical channels}, the contact potential of Eq.~\eqref{eq:V_contact_OPE} for $X(3872)$ regarded as a molecular $[\bar{D}D^*]$ state can be expressed in terms of the isoscalar and isovector couplings with Eq.~\eqref{eq:X physical and isospin channels} as
\begin{eqnarray}
V_{C, X}=\frac{1}{2}
\begin{pmatrix}
 C_{0X}+C_{1X} & C_{0X}-C_{1X} \\
 C_{0X}-C_{1X} & C_{0X}+C_{1X}
\end{pmatrix} ,
\label{eq:LECs in EFT of X}
\end{eqnarray}
where $C_{0X}$ and $C_{1X}$ denote the contact interactions in the isoscalar and isovector channels, respectively.

Analogously, the contact potential in the physical basis for $T_{cc}(3875)^+$ regarded as a molecular $[DD^*]$ state is
\begin{eqnarray}
V_{C, T}=\frac{1}{2}
\begin{pmatrix}
 C_{0T}+C_{1T} & C_{1T}-C_{0T} \\
 C_{1T}-C_{0T} & C_{0T}+C_{1T}
\end{pmatrix} .
\label{eq:LECs in EFT of Tcc}
\end{eqnarray}

These LECs of $C_{0X}$, $C_{1X}$, $C_{0T}$, and $C_{1T}$ are independent parameters unless additional dynamical assumptions are imposed: heavy-meson EFT alone does not identify the meson-antimeson and meson-meson contact interactions. To test a common molecular description of the $X(3872)$ and $T_{cc}(3875)^+$, we relate these couplings through a meson-exchange saturation model.

Here we employ the strategy of Refs.~\cite{Peng:2020xrf,Peng:2021hkr,Yang:2022ezl,Peng:2023lfw}, in which these LECs of the contact-range potential are saturated by the exchange of the scalar $\sigma$ and vector $\rho$, $\omega$ mesons, combined with the fact that they are running couplings governed by the renormalization group (RG) of the effective field theory.

First, the leading-order effective Lagrangians for the interactions between the heavy mesons and the exchanged light mesons are given, for simplicity, in terms of the light-quark degrees of freedom $q_L$ as
\begin{eqnarray}
\label{eq:Lagrangian for light meson exchange}
\mathcal{L}_{S}^0 &=& g_{S} \,q_{L}^{\dagger}\sigma q_L \, , \nonumber
\\
\mathcal{L}_{\rho}^0
&=&
g_{\rho}\,
q_L^\dagger
\bm{\tau}^a
\rho_0^a
q_L
-
\frac{f_{\rho}}{2M}\,
q_L^\dagger
\bm{\tau}^a
\epsilon_{ijk}
\vec{\sigma}_{Li}
\partial_j \bm{\rho}_k^a
q_L
+\cdots \, ,  \nonumber
\\
\mathcal{L}_{\omega}^0
&=&
-g_{\omega}\,
q_L^\dagger
\omega_0
q_L
+
\frac{f_{\omega}}{2M}\,
q_L^\dagger
\epsilon_{ijk}
\vec{\sigma}_{Li}
\partial_j \bm{\omega}_k
q_L
+\cdots \,.\nonumber
\end{eqnarray}
Here $\vec{\sigma}_L$ denotes the Pauli matrices acting on the spin of the light quark inside the heavy hadron, and the dots indicate higher-order multipole terms. We use $M=m_N\approx 938.9 \, \rm MeV$, $g_S=3.4$, $g_V=g_\rho=g_\omega=2.9$, and $f_V=f_\rho=f_\omega=k_V g_V$, with $k_V=2.85$, following Ref.~\cite{Peng:2021hkr,Peng:2023lfw}. From the above Lagrangians, the exchange potential from the scalar meson $\sigma$ is
\begin{eqnarray}
    V_S(\bm q)=-\frac{g_S^2}{\bm q^2+m_{\sigma}^2} \, .
    \label{eq:potential of sigma}
\end{eqnarray}

For the heavier vector-meson exchanges, the $S$-wave pieces of the vector-meson exchange potentials read
\begin{eqnarray}
V_{\rho}(\bm q) &=& \tau_{\rm fac}\,\left( \frac{g_{\rho}^2}{\bm q^2+m_{\rho}^2} - \frac{\eta \, S_{\rm fac} f_{\rho}^2}{6M^2}\frac{\bm q^2}{\bm q^2+\mu_{\rho}^2} \right) \, , \label{eq:potential of rho} \\
V_{\omega}(\bm q) &=& -\zeta \,\left( \frac{g_{\omega}^2}{\bm q^2+m_{\omega}^2} -  \frac{\eta \, S_{\rm fac} f_{\omega}^2}{6M^2}\frac{\bm q^2}{\bm q^2+\mu_{\omega}^2} \right) \, ,
\label{eq:potential of omega}
\end{eqnarray}
where $\mu_{\rho,\omega}^2=m_{\rho,\omega}^2-\Delta_D^2$ with $m_\rho\simeq 770~{\rm MeV}$ and $m_\omega\simeq 783~{\rm MeV}$, and the values of the spin factor $S_{\rm fac}=\braket{\vec{\sigma}_{1}\cdot\vec{\sigma}_{2}}$ are listed in Table~\ref{Table:lambda values}. The magnetic-dipole terms in Eqs.~\eqref{eq:potential of rho} and \eqref{eq:potential of omega} also contain local contributions. Neglecting $\Delta_D^2$ relative to the vector-meson masses squared, their momentum dependence can be decomposed as
\begin{equation}
\frac{\bm q^2}{\bm q^2+m_{\rho/\omega}^2} = 1- \frac{m_{\rho/\omega}^2}{\bm q^2+m_{\rho/\omega}^2} \, ,
\label{eq:decomposition for vector M1}
\end{equation}
where the first term is a momentum-independent constant whose Fourier transform is the local $\delta^{(3)}(\bm r)$ contribution. However, we do not apply the same subtraction prescription as for the pion exchange to the $\rho$- and $\omega$-exchange contributions, because the vector-meson masses are of the same order as the ultraviolet scale $\Lambda$ used to regulate the low-energy interaction, $m_{\rho,\omega}\sim\Lambda$. Consequently, $1/m_{\rho,\omega} \sim 1/\Lambda$, and there is no parametric separation between a regulator-localized contribution of range $1/\Lambda$ and the intrinsic range of the vector-meson exchange. In contrast to the pion, the $\rho$ and $\omega$ exchanges therefore do not generate a long-range tail at a distance scale that is parametrically separated from the ultraviolet scale.

Accordingly, the $\rho$- and $\omega$-exchange contributions are regarded as part of the short-distance dynamics and are incorporated into the contact interaction through resonance saturation. We do not remove their local $\delta^{(3)}(\bm r)$ components separately. To study the impact of this local potential on molecular interaction for exotic hadrons, we replace the constant $1$ in Eq.~\eqref{eq:decomposition for vector M1} by $\delta_a\in[0,1]$. The endpoints $\delta_a=0$ and $\delta_a=1$ correspond to omitting and retaining that local term, respectively.

The LECs appearing in the effective-field-theory contact potentials of Eqs.~\eqref{eq:LECs in EFT of X} and \eqref{eq:LECs in EFT of Tcc} can now be related to each other by applying resonance saturation with the scalar $\sigma$ and vector $\rho$, $\omega$ exchanges of Eqs.~\eqref{eq:potential of sigma}, \eqref{eq:potential of rho}, and \eqref{eq:potential of omega} as
\begin{eqnarray}
     C_{\sigma}^{sat} (\mu\sim m_{\sigma}) \!\! &\propto& \!\! -\frac{g_S^2}{m_{\sigma}^2} \,,
     \nonumber \\
     C_{\rho}^{sat} (\mu\sim m_{V}) \!\! &\propto& \!\! \tau_{\rm fac}\,\left[ \frac{g_{V}^2}{m_{V}^2} -  \frac{\eta \, S_{\rm fac} f_{V}^2}{6M^2}\left( \delta_a-1\right)\right]  ,
      \\
     C_{\omega}^{sat} (\mu\sim m_{V}) \!\! &\propto& \!\! -\zeta \,\left[ \frac{g_{V}^2}{m_{V}^2} -  \frac{\eta \, S_{\rm fac} f_{V}^2}{6M^2}\left( \delta_a-1\right)\right] \,, \nonumber
\end{eqnarray}
where, for simplicity, we neglect the small differences between the $\rho$ and $\omega$ parameters, set $m_V=(m_\rho+m_\omega)/2$, and use $\mu_{\rho,\omega}^2\simeq m_{\rho,\omega}^2$. At $\delta_a=1$, the local and finite-range contributions from magnetic dipole cancel at zero momentum in this expansion.

We do not simply take the LECs from resonance saturation as constants; we use proportional relations rather than equalities, because the LECs in effective field theory depend on the ultraviolet scale $\Lambda$ introduced in the Lippmann-Schwinger equation and therefore run with that scale. Resonance saturation works best at a regularization scale $\mu$ of the same order as the mass of the exchanged particle, $\mu\sim m_{\rm ex}$, so the above proportional relations define these LECs at their natural scales. To combine these relations at a common scale, the renormalization group of our effective field theory is used to run the LECs to a single, freely chosen scale; here we present the combination at the representative scale $\mu\sim m_V$. We adopt the short-distance scaling prescription following Refs.~\cite{PavonValderrama:2014zeq,Valderrama:2016koj,Yan:2026zvw}, in which the contact contribution satisfies the renormalization group equation
\begin{eqnarray}
    \frac{d}{d\Lambda}\bra{\Psi(R)} C(\Lambda)\ket{\Psi(R)}=0 \, ,
\end{eqnarray}
where $\Lambda$ is the ultraviolet cutoff and the wave function of $\Psi (R)\sim \Lambda^{-1/2}$ at distances $R\sim 1/\Lambda$. Thus, we have the scaling relation
\begin{eqnarray}
C(\Lambda_1)=\frac{\Lambda_1}{\Lambda_2}\,C(\Lambda_2) \, .
\end{eqnarray}
The saturated couplings from the scalar $\sigma$ and vector $\rho$, $\omega$ exchanges can then be run to the common renormalization scale $\mu\sim m_V$ and combined into
\begin{eqnarray}
&& C^{\rm sat}(\mu\sim m_V) =  C_{\rho}^{sat}(\mu\sim m_V)+ C_{\omega}^{sat}(\mu\sim m_V) \nonumber \\
    && + C_{\sigma}^{sat}(\mu\sim m_V)  \propto \left(\tau_{\rm fac}-\zeta\right) \times  \nonumber \\
    &&  \left[ \frac{g_{V}^2}{m_{V}^2} -  \frac{\eta \, S_{\rm fac} f_{V}^2}{6M^2}\left( \delta_a-1\right)\right] -\frac{m_V}{m_\sigma}\frac{g_S^2}{m_\sigma^2} \, .
    \label{eq:saturated LECs at same scale}
\end{eqnarray}
Within this common scaling prescription, the ratios of the saturated couplings are unchanged when they are evolved to the cutoff used in the Lippmann--Schwinger equation. Eq.~\eqref{eq:saturated LECs at same scale}, together with Table~\ref{Table:lambda values}, therefore fixes the relations among $C_{0X}$, $C_{1X}$, $C_{0T}$, and $C_{1T}$. For each choice of $\delta_a$, $m_\sigma$, and $\Lambda$, the overall strength is subsequently calibrated using the $T_{cc}(3875)^+$ input, as described in Sec.~\ref{sec4: numerical results}.

The resulting interaction kernels from Eqs.~\eqref{eq: sub-OPE for X and T}-\eqref{eq:LECs in EFT of Tcc} obtained for $X(3872)$ and $T_{cc}(3875)^+$ therefore read
\begin{eqnarray}
\begin{aligned}
    V_{X}(E;\bm p', \bm p)=V_{C,X}+V_{\mathrm{OPE}, \, X}^{\rm sub}(E;\bm p', \bm p) \,,
    \\
    V_{T}(E;\bm p', \bm p)=V_{C,T}+V_{\mathrm{OPE}, \, T}^{\rm sub}(E;\bm p', \bm p) \,\,.
\end{aligned}
    \label{eq:VX VT potential}
\end{eqnarray}

\section{Scattering equation and analytic continuation}
\label{sec3: LS equation and sigularities}

\subsection{Lippmann-Schwinger equation and the unstable $D^*$ self-energy}

\begin{figure}[htbp]
    \centering
\includegraphics[width=\linewidth]{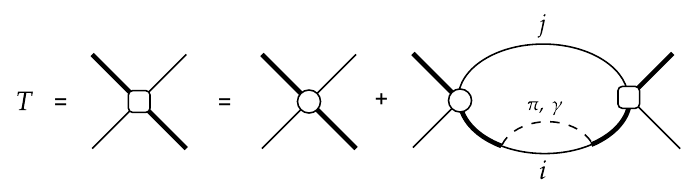}
    \caption{Diagrammatic representation of the Lippmann-Schwinger equation with the $D^*$ self-energy included in the two-body propagator. The square and the circle denote the $T$-matrix amplitude and the interaction kernel $V(E;\bm p', \bm k)$ in Eq.~\eqref{eq:LS equation}, respectively. Thick (thin) solid lines represent vector (pseudoscalar) charmed mesons, and dashed lines represent the pion or photon in the $D^*$ self-energy.}     \label{fig:LSE plot}
\end{figure}

With the potentials of Eq.~\eqref{eq:VX VT potential} in the molecular basis of Eqs.~\eqref{eq:X physical channels} and \eqref{eq:T physical channels}, the Lippmann-Schwinger equation (LSE) can be formulated to determine the coupled-channel scattering amplitude. As shown in Fig.~\ref{fig:LSE plot}, the LSE reads
\begin{eqnarray}
&&T(E;\bm p',\bm p)=V(E;\bm p', \bm p) + \nonumber \\
&& \int_0^\Lambda \frac{d^3\bm k}{(2\pi)^3}\,V(E;\bm p', \bm k)\,G (E, k)\,T(E;\bm k,\bm p),
    \label{eq:LS equation}
\end{eqnarray}
where the kernel $V(E;\bm p', \bm p)$ denotes the potentials of Eq.~\eqref{eq:VX VT potential}, the diagonal matrix $G(E,k)$ is the nonrelativistic two-body propagator, and the integration is regulated by a sharp momentum cutoff $|\bm k|<\Lambda$. For the $S$-wave kernels, the angular integration reduces the measure to $k^2dk/(2\pi^2)$. To consider the full three-body effects, the self-energy of the unstable $D^*$ must be incorporated into the propagator as
\begin{eqnarray}
 G_{11}(E, k) &=& \frac{1}{E-m_1-m_1^*-\frac{k^2}{2\mu_1}+i\frac{\Gamma_1(E,k)}{2}} \, , \label{eq:G propagator of channel 1}
 \\
G_{22}(E, k) &=& \frac{1}{E-m_2-m_2^*-\frac{k^2}{2\mu_2}+i\frac{\Gamma_2(E,k)}{2}} \, , \label{eq:G propagator of channel 2}
\end{eqnarray}
where $\mu_i=m_im_i^*/(m_i+m_i^*)$ is the reduced mass in channel $i=1,2$, while $\Gamma_{i}(E,k)$ the $D^*$ decay widths, which consist of the radiative decay and the strong decay into $D\pi$. For $D^{*+}$, the radiative contribution is estimated from the total width and branching fraction quoted by the Particle Data Group~\cite{ParticleDataGroup:2026mpi}:
\begin{eqnarray}
 \Gamma_{D^{*+}}^{R}&=83.4\,\, \rm keV\times 1.6\%\approx1.33 \,\rm keV \, ,
\end{eqnarray}
while for the neutral $D^{*0}$, the radiative decay width is obtained from the predicted total width and the measured branching ratio as~\cite{Rosner:2013sha,Guo:2019qcn,Braaten:2020iye,ParticleDataGroup:2026mpi}
\begin{eqnarray}
 \Gamma_{D^{*0}}^{R}&=55.3 \,\, \rm keV \times 35.3\%\approx19.52 \,\,\rm keV \, .
\end{eqnarray}

For the strong decay $D^*\rightarrow D+\pi$ entering the self-energy, which is a $P$-wave two-body process, the decay widths for the neutral and charged $D^*$ mesons read~\cite{Baru:2011rs,Du:2021zzh,Zhang:2024fxy}
\begin{widetext}
\begin{eqnarray}
    \Gamma^{\pi}_{D^{*0}} (E,k)&=&
    \frac{g_{\pi}^2 \,m_{D^0}}{24\pi f_{\pi}^2 \,m_{D^{*0}}} P^{3}_{D^0\pi^0 j}(E,k) 
    +\frac{g_{\pi}^2 \,m_{D^{\pm}}}{12\pi f_{\pi}^2 \,m_{D^{*0}}}\left[ P^{3}_{D^{\pm}\pi^{\mp} j}(E,k)-  P^{3}_{D^{\pm}\pi^{\mp} j}(m_{j}+m_{D^{*0}},0)\right] \, ,
    \label{eq:width of Dstar 0}
    \\
    \Gamma^{\pi}_{D^{*\pm}} (E,k)&=&
    \frac{g_{\pi}^2 \,m_{D^{\pm}}}{24\pi f_{\pi}^2 \,m_{D^{*\pm}}} P^{3}_{D^{\pm}\pi^0 j}(E,k)
    +\frac{g_{\pi}^2 \,m_{D^0}}{12\pi f_{\pi}^2 \,m_{D^{*\pm}}} P^{3}_{D^0\pi^\pm j}(E,k) \, .
    \label{eq:width of Dstar c}
\end{eqnarray}
\end{widetext}
Here, $P_{i\pi j}$ is the center-of-mass momentum of the two-body $D\pi$ system. In the nonrelativistic approximation,
\begin{eqnarray}
    P_{i\pi j}(E,k)=\sqrt{2\mu_{i\pi}\left(E'-m_i-m_\pi\right)} \,,
    \label{eq:CM momentum of D-pi}
\end{eqnarray}
where $E'$ is the center-of-mass energy of the unstable $D^*/\bar{D}^*$ meson (with mass denoted by $m^*$) inside the propagator, which can be estimated as
\begin{eqnarray}
   E'=E+m^{\ast}-\sqrt{m^{\ast 2}+k^2}-\sqrt{m_j^2+k^2} \, ,
    \label{eq:static energy of E'}
\end{eqnarray}
where $\mu_{i\pi}$ is the reduced mass of the two-body $D\pi$ system, and $i$, $j$, and $\pi$ label the two $D$ mesons and the pion shown in Fig.~\ref{fig:LSE plot}.

For the neutral $D^{*0}$ meson, the $D^{\pm}\pi^{\mp}$ channel is closed at its physical mass, but the corresponding virtual pion loop generates a dispersive mass correction. Since the propagator is written in terms of the physical $D^{*0}$, the constant on-shell contribution of this loop is subtracted to avoid shifting the input mass. Specifically, we subtract the loop function evaluated at $E=m_j+m_{D^{*0}}$ and $k=0$, where $m_j$ is the spectator mass of meson $j$ in Fig.~\ref{fig:LSE plot}, so that the energy available to the $D^{*0}$ is $E'=m_{D^{*0}}$. This reduces to $E=m_{D^+}+m_{D^{*0}}$ in the $D^+D^{*0}$ channel for $T_{cc}(3875)^+$ considered in Ref.~\cite{Du:2021zzh}. The subtraction absorbs the constant mass correction while retaining the energy dependence and threshold structure of the loop. For the charged $D^{*\pm}$ mesons, both strong-decay channels are open at the physical mass, and the retained $P$-wave loop terms contribute to the on-shell width without an analogous finite subtraction.

\subsection{Singularities and integration contours}

The pole search requires analytic continuation of both the propagators and the OPE kernel. Several kinds of singularity structures emerge when solving the LSE in Eq.~\eqref{eq:LS equation}, which are introduced below.
\begin{itemize}
    \item Two-body channel structure: each of the molecular channels in Eqs.~\eqref{eq:X physical channels} and \eqref{eq:T physical channels} introduces a branch point of the on-shell momentum $k_{\rm on}=\sqrt{2\mu\left( E-m-m^*+i\frac{\Gamma(E,k)}{2} \right)}$ in the denominator of the propagator $G(E,k)$ in Eqs.~\eqref{eq:G propagator of channel 1} and \eqref{eq:G propagator of channel 2}. Since the self-energy $\Gamma(E,k)$ of the unstable $D^*$ is included in the propagator, the energy cut starts naturally at the two-body threshold and extends to infinity with an imaginary part. Therefore, the different branches of $k_{\rm on}$ introduce four Riemann sheets for the coupled-channel system in Eqs.~\eqref{eq:X physical channels} and \eqref{eq:T physical channels}. When solving the $T$-matrix, the bound-state poles are located on the first (physical) Riemann sheet, while other kinds of poles lie on unphysical sheets.

    \item Three-body cut from the self-energy of $D^*/\bar{D}^*$: the branch point in Eq.~\eqref{eq:CM momentum of D-pi} introduces a cut, which is actually associated with the on-shell three-body cut starting from $m_i+m_\pi+m_j$ (see Fig.~\ref{fig:LSE plot}), with the nonrelativistic expansion of Eq.~\eqref{eq:static energy of E'}, $E'=E-m_j-k^2/(2\mu)$, where $\mu$ is the reduced mass of the spectator and the unstable meson.

    When searching for the $T$-matrix poles, this cut would obstruct the analytic continuation of the pole search from the upper to the lower complex energy plane. Therefore, we redirect the cut of $E'$ along the imaginary axis to avoid this problem, by redefining Eq.~\eqref{eq:CM momentum of D-pi} as
\begin{eqnarray}
    P_{i\pi j}(E,k) = -\sqrt{2\mu_{i\pi}\left(E'-m_i-m_\pi\right)} ,     \label{eq: redirected cut 1}
\end{eqnarray}
for $\mathrm{Re}(E')>m_i+m_\pi$ and $\mathrm{Im}(E')<0$. Otherwise,
\begin{eqnarray}
    P_{i\pi j}(E,k) = \sqrt{2\mu_{i\pi}\left(E'-m_i-m_\pi\right)} .    \label{eq: redirected cut 2}
\end{eqnarray}

    Meanwhile, with the real integration path $k\in[0,\,\Lambda]$ in Eq.~\eqref{eq:LS equation}, the static $D^*$ energy $E'$ can still cross the redirected cut in Eqs.~\eqref{eq: redirected cut 1} and \eqref{eq: redirected cut 2} when the $E$ poles are searched for in the lower complex plane with $\mathrm{Im}(E)<0$. Therefore, a distorted integration path, the spectator momentum contour (SMC), is introduced in the form~\cite{Doring:2009yv}
\begin{eqnarray}
    k_{\mathcal C}=k_r+iV_0\left(1-e^{-k_r/w}\right)\left(1-e^{(k_r-\Lambda)/w}\right), 
\end{eqnarray}
with $k_r\in[0,\Lambda]$. Choosing $V_0<0$ and $w>0$ deforms the contour into the lower half of the momentum plane while keeping its endpoints fixed. The deformation must avoid the singularities of the integrand and preserve the chosen continuation.

    \item Singularities of the OPE potential kernel: the angular integral in Eq.~\eqref{eq: OPE potential partial waves} is sensitive to the zeros and branch points associated with $E_\pi(\bm q)$ and $D_{\alpha\beta\pi}$. The former arise from the exchanged-pion energy, while the latter include the forward three-body intermediate state. The backward cut $C_2$ cannot go on shell at the energies near thresholds, so $D_{\alpha^*\beta^*\pi}$ contributes no singularity here.

    Following Refs.~\cite{Dong:2024fjk,Zhang:2024fxy}, we deform the $z$ integration contour according to the positions of the relevant singularities. The spectator and angular contours must be chosen consistently when continuing the amplitude to search for a pole.
\end{itemize}

\subsection{Isospin breaking in the $X(3872)$ decays}
The isospin-symmetry violation in the decays of $X(3872)$ to the isoscalar $J/\psi\,\omega\,(\pi^+\pi^-\pi^0)$ and isovector $J/\psi\,\rho^0\,(\pi^+\pi^-)$ final states is an important feature of the $X(3872)$ state~\cite{Wang:2022xga}. The relation between the physical $[\bar{D}D^*]$ channels and the isospin channels in Eq.~\eqref{eq:X physical and isospin channels} provides a convenient way to estimate this isospin-symmetry violation. The ratio of the $X(3872)$ couplings to $J/\psi\rho^0$ and $J/\psi\omega$ has been updated using the Omn\`{e}s parametrization~\cite{Dias:2024zfh} as
\begin{eqnarray}
    R_X=\frac{g_{X\psi\rho}}{g_{X\psi\omega}}=0.26\pm 0.03 \, .
    \label{eq: RX isospin violation of X3872}
\end{eqnarray}
In our framework, the couplings $g_{XN}$ and $g_{XC}$ to the neutral and charged molecular channels can be extracted from the residues of the $T$-matrix at the pole. Assuming equal short-distance transition strengths for the isoscalar and isovector $\bar{D}D^*\to J/\psi \rho(\omega)$ amplitudes, within the molecular picture, the ratio $R_X$ is estimated as~\cite{Hidalgo-Duque:2012rqv,Zhang:2024fxy}
\begin{eqnarray}
    R_X=\left|\frac{g_{XN}-g_{XC}}{g_{XN}+g_{XC}}\right| \, .
    \label{eq: predicted RX isospin violation of X3872}
\end{eqnarray}

\section{Numerical results}
\label{sec4: numerical results}

To understand the short-range dynamics, especially the potential saturated by the $\delta^{(3)}(\bm r)$ magnetic-dipole term of the vector-meson exchange potentials in Eqs.~\eqref{eq:potential of rho} and \eqref{eq:potential of omega}, we vary the parameter $\delta_a$ in Eq.~\eqref{eq:saturated LECs at same scale} from $0$ to $1$ in steps of $0.1$. At each $\delta_a$ point, the real part of the $T_{cc}(3875)^+$ pole fixes the contact couplings, and the saturation relations then determine the interaction in the $X(3872)$ sector. This procedure tests the correlated description of the two states within the adopted saturation model. Resonance saturation is expected to be most reliable when the cutoff is comparable to the masses of the exchanged mesons. We use $\Lambda=0.8$, $1.0$, and $1.2\,\mathrm{GeV}$, which bracket the vector-meson mass scale $m_V\simeq780\,\mathrm{MeV}$ and allow us to assess the cutoff dependence. We also vary the effective scalar-exchange mass over $m_\sigma=400$, $475$, and $550\,\mathrm{MeV}$. This variation probes the sensitivity to the poorly constrained scalar meson mass.

In addition, to assess the effect of explicit pion exchange, we compare two interaction schemes:
\begin{itemize}
\item[(i)] Scheme $A$ contains only the contact potential in the LSE kernel. The $D^*/\bar D^*$ self-energy is still included in the propagator $G(E,k)$, so partial three-body effects are retained in this scheme.
\item[(ii)] Scheme $B$ contains both the contact potential and the subtracted OPE potential. Together with the $D^*/\bar D^*$ self-energy, it accounts for the full three-body effects.
\end{itemize}
In scheme $B$, OPE is iterated in the LSE. Comparing the two schemes measures its net effect after the contact strength has been determined separately from the same $T_{cc}(3875)^+$ input.

\subsection{Connections between the contact interactions of $X(3872)$ and $T_{cc}(3875)^+$}

The saturation prescription in Eq.~\eqref{eq:saturated LECs at same scale} gives the dimensionless ratios
\begin{eqnarray}
CR_X&=&\frac{C_{1X}}{C_{0X}},
\,\,
CR_T=\frac{C_{1T}}{C_{0T}},
\,\,
CR_{XT}=\frac{C_{0X}}{C_{0T}}.
\label{eq: contact couplings connection between X and T}
\end{eqnarray}
Here, $C_{0X}$ and $C_{1X}$ are the isoscalar and isovector contact couplings in the $\bar{D} D^*$ sector for $X(3872)$, and $C_{0T}$ and $C_{1T}$ are their counterparts in the $DD^*$ sector for $T_{cc}(3875)^+$. The common scale evolution adopted in the saturation prescription preserves these ratios. They determine the correlation between the two sectors once the overall contact strength is fixed.

The $T_{cc}(3875)^+$ pole extracted from the line shape, expressed relative to the $D^0D^{*+}$ threshold as $E_{p;T}=\sqrt{s_{p;T_{cc}^+}}-m_{D^0}-m_{D^{*+}}$, is~\cite{LHCb:2021auc}
\begin{eqnarray}
E_{p;T} = (-360\pm40^{+4}_{-0}) -i(24\pm1^{+0}_{-7})~\mathrm{keV}.
\label{eq:Tcc pole from experiments}
\end{eqnarray}
Taking the central real part of this pole position, $-360~{\rm keV}$, to fix the overall strength of the contact potential. The ratio $CR_T$ gives $C_{1T}$ relative to $C_{0T}$, so one real input suffices within this parametrization. The contact couplings and the pole are therefore determined iteratively, where the imaginary part of the pole is a prediction of the included decay dynamics.

For $T_{cc}(3875)^+$, the real part of the converged pole remains approximately $-360\,\mathrm{keV}$ throughout the parameter scan. Scheme $A$ gives $\Gamma_{T_{cc}^+}/2\simeq17\,\mathrm{keV}$. In scheme $B$, the half-width is approximately $28\,\mathrm{keV}$ for $\Lambda=0.8$ and $1.2\,\mathrm{GeV}$, and $25\,\mathrm{keV}$ for $\Lambda=1.0\,\mathrm{GeV}$. Scheme $B$, particularly at $\Lambda=1.0\,\mathrm{GeV}$, gives a half-width close to the experimental central value of $24\,\mathrm{keV}$. This decay enhancement relative to scheme $A$ agrees with the trend found in previous calculations of pion-exchange and three-body effects~\cite{Du:2021zzh,Qiu:2023uno}.

\begin{figure}[htbp]
    \centering
\includegraphics[width=1\linewidth]{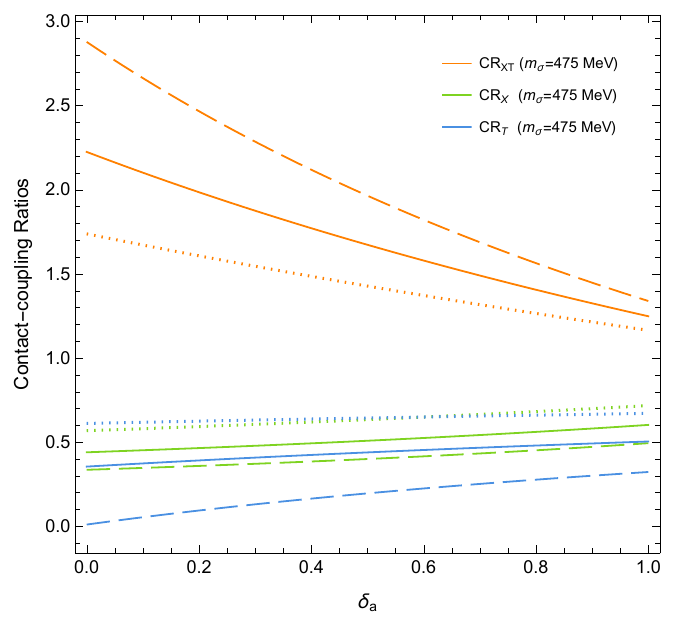}
\caption{Contact coupling ratios as functions of $\delta_a$. Orange, green, and blue curves show $CR_{XT}$, $CR_X$, and $CR_T$, respectively. Dotted, solid, and dashed curves correspond to $m_\sigma=400$, $475$, and $550\,\mathrm{MeV}$. The legend labels the solid curves at the reference value $m_\sigma=475\,\mathrm{MeV}$.}    \label{fig:Ratios changing}
\end{figure}

Once the $T_{cc}(3875)^+$ contact strength is fixed, $CR_{XT}$ and $CR_X$ determine $C_{0X}$ and $C_{1X}$. The $X(3872)$ pole is thus a prediction of the correlated saturation model. Fig.~\ref{fig:Ratios changing} shows that $CR_X$ and $CR_T$ vary more slowly with $\delta_a$ than $CR_{XT}$. For the attractive isoscalar couplings considered here, the decrease in $CR_{XT}$ reduces the relative attraction in the $X(3872)$ sector. As a result, increasing $\delta_a$ while keeping the $T_{cc}(3875)^+$ pole fixed generally produces a shallower $\bar{D}D^*$ state.

\subsection{Pole trajectories and isospin violation of the predicted $X(3872)$} \label{subsec: Pole trajectories}

\begin{figure}[htbp]
    \centering
\includegraphics[width=\linewidth]{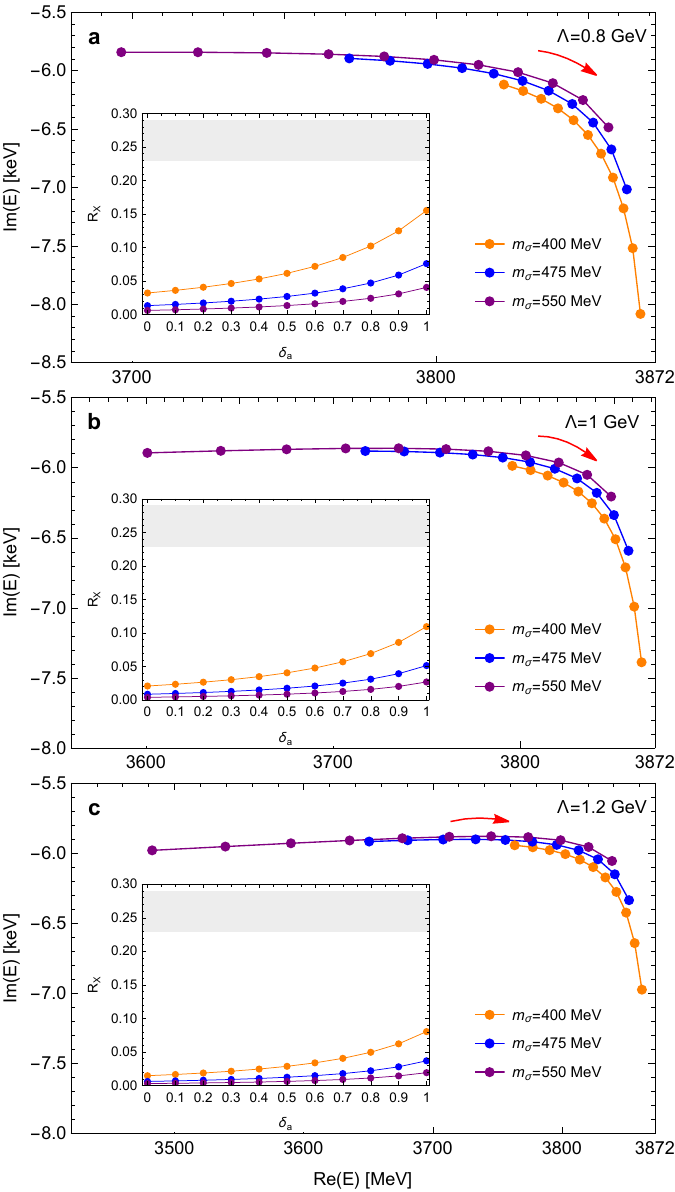}
    \caption{Predicted $[\bar D D^*]$ pole trajectories in scheme $A$, with the contact interaction fixed by the $T_{cc}(3875)^+$ input. Panels $\rm a$--$\rm c$ correspond to $\Lambda=0.8$, $1.0$, and $1.2\,\mathrm{GeV}$. Orange, blue, and purple curves denote $m_\sigma=400$, $475$, and $550\,\mathrm{MeV}$, respectively. Successive points correspond to $\delta_a=0,0.1,\ldots,1$, and the arrows indicate increasing $\delta_a$. The main panels show the pole energies, with real and imaginary parts in MeV and keV, respectively. The insets show the residue-based ratio $R_X$, with the shaded band indicating $0.26\pm0.03$, as given in Eq.~\eqref{eq: RX isospin violation of X3872}.}
    \label{fig:Pole trajectory with only Contact}
\end{figure}

\begin{figure}[htbp]
    \centering
\includegraphics[width=1\linewidth]{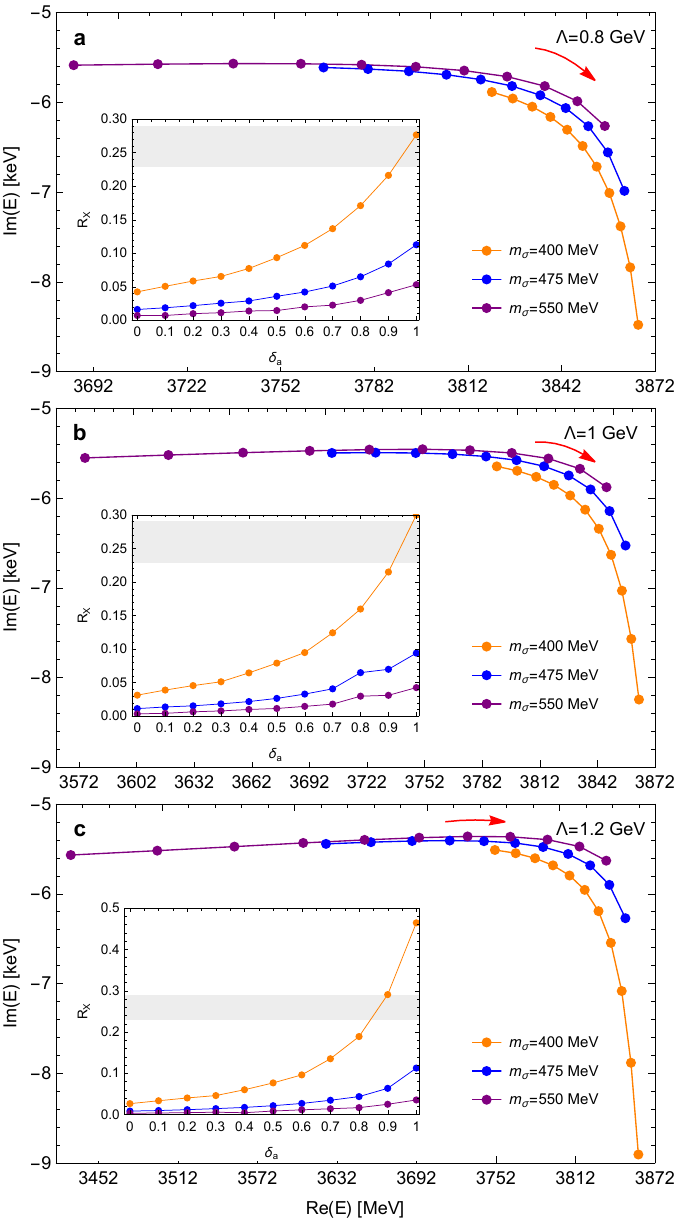}
    \caption{As in Fig.~\ref{fig:Pole trajectory with only Contact}, but for scheme $B$, which includes the subtracted OPE potential in addition to the contact interaction. The contact strength is determined separately from the same $T_{cc}(3875)^+$ pole input.}
    \label{fig:Pole trajectory with OPE}
\end{figure}

Figs.~\ref{fig:Pole trajectory with only Contact} and \ref{fig:Pole trajectory with OPE} show the predicted $[\bar D D^*]$ pole
trajectories in schemes $A$ and $B$, respectively. The poles are represented as
$E=M-i\Gamma/2$, with the real part corresponding to the mass and the
imaginary part to the half decay width. Increasing $\delta_a$ moves the poles toward the neutral threshold, with the closest approach occurring at $\delta_a=1$ in the parameter range investigated. The widths also change along the trajectories, particularly near their shallowest endpoints. At fixed $m_\sigma$ and $\delta_a$, a larger cutoff generally produces a more deeply bound state. This remaining cutoff dependence is relevant when judging the accuracy of the correlated prediction.

The insets of Figs.~\ref{fig:Pole trajectory with only Contact} and \ref{fig:Pole trajectory with OPE} display $R_X$, calculated from the pole residues using Eq.~\eqref{eq: predicted RX isospin violation of X3872}. Scheme $B$ gives larger values than scheme $ A$ systematically. Within the scan, scheme $A$ does not reach the reference band $R_X=0.26\pm0.03$, whereas some scheme-$B$ parameter points do. Therefore, although the OPE interaction has only a modest effect on the pole position as discussed below, it plays an essential role in reproducing the isospin-violating properties of $X(3872)$. This distinction is important: a subleading contribution to the binding energy or resonance width does not necessarily imply a negligible effect on observables that are particularly sensitive to the relative weights of different isospin components.

The endpoint closest to the observed $X(3872)$ mass in scheme $B$ occurs at $m_\sigma=400\,\mathrm{MeV}$ and $\delta_a=1$. The corresponding real pole energies are $3866.63$, $3863.67$, and $3859.48\,\mathrm{MeV}$ for $\Lambda=0.8$, $1.0$, and $1.2\,\mathrm{GeV}$, respectively. These values remain approximately $5$--$12\,\mathrm{MeV}$ below the near-threshold pole extracted from experiments. Retaining the local vector-exchange contribution reduces the mass discrepancy. The shallowest endpoint remains the overbound $[\bar{D}D^*]$ compared to the $X(3872)$ pole, identifying a tension in the correlated molecular picture for $T_{cc}(3875)^+$ and $X(3872)$.

\begin{figure}[htbp]
    \centering
\includegraphics[width=1\linewidth]{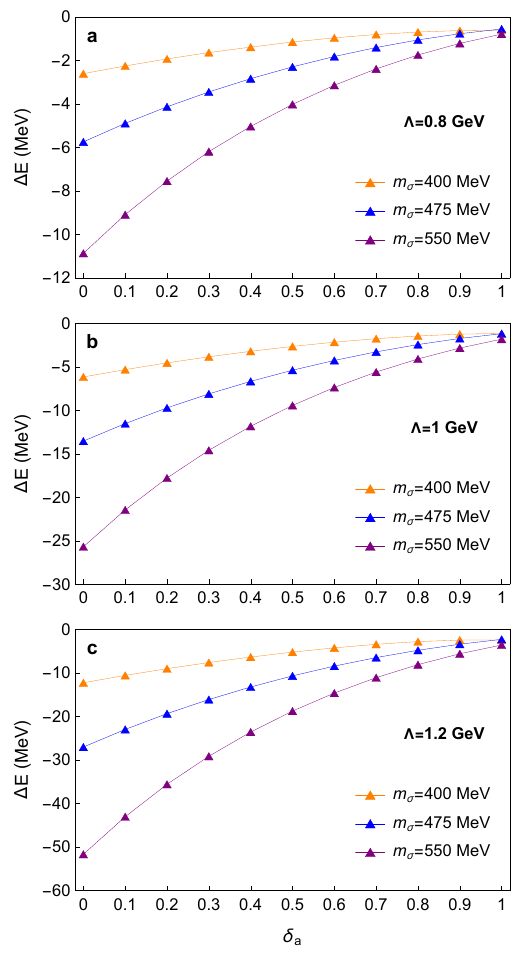}
 \caption{Difference $\Delta E$ between the real parts of the predicted $[\bar{D}D^*]$ molecular poles obtained in schemes $A$ and $B$, as a function of $\delta_a$. Panels $\rm a$--$\rm c$ correspond to the three cutoffs $\Lambda=0.8\,\rm GeV$, $1\,\rm GeV$, and $1.2\,\rm GeV$, respectively. In each panel, the results for the three choices of $m_\sigma$ considered for the uncertainty are plotted in orange, blue, and purple.}
    \label{fig:Pole differences with OPE and Contact}
\end{figure}

Comparing the two schemes $A$ and $B$, we define 
\begin{equation}
\Delta E=\mathrm{Re}\,E^{B}-\mathrm{Re}\,E^{A}
\end{equation}
to measure the difference between the real parts of the predicted poles obtained in the two schemes. As shown in Fig.~\ref{fig:Pole differences with OPE and Contact}, $\Delta E$ is negative throughout the scan, meaning scheme $B$ predicts a more deeply bound $[\bar D D^*]$ pole. This behavior is expected: the pion-exchange potential provides additional attraction and makes the predicted states more deeply bound in scheme $B$. The magnitude of $\Delta E$ decreases as $\delta_a$ increases and is largest where the pole predicted in scheme $A$ is most deeply bound.

\begin{figure}[htbp]
    \centering
\includegraphics[width=1\linewidth]{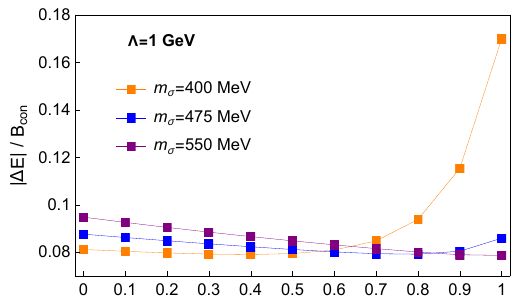}
    \caption{Ratio $|\Delta E|/\mathrm{B_{con}}$ as a function of $\delta_a$, where $\mathrm{B_{con}}$ is the binding energy of the $[\bar{D}D^*]$ system predicted within scheme $A$. Only the cutoff $\Lambda=1\,\rm GeV$ is shown here; the other cutoffs show the same qualitative behavior of this ratio.}
    \label{fig:Pole differences ratio with OPE and Contact}
\end{figure}

To quantify the correction induced by the OPE potential, we define the ratio $|\Delta E|/\mathrm{B_{con}}$, where $\mathrm{B_{con}}=\mathrm{m}_{\bar{D}^0}+\mathrm{m}_{D^{*0}}-\mathrm{Re}\,E^{A}$ is the binding energy obtained in scheme $A$ with the pure contact potential. As shown in Fig.~\ref{fig:Pole differences ratio with OPE and Contact} for $\Lambda=1 \,\rm GeV$ (the other cutoffs show the same qualitative behavior), $|\Delta E|/\mathrm{B_{con}}$ is approximately $0.08$--$0.10$ for $m_{\sigma}=475$ and $550 \,\rm MeV$. This indicates that the subtracted OPE potential $V_{\mathrm{OPE}, \, X}^{\rm sub}(E;\bm p', \bm p)$ provides, to a good approximation, a perturbative correction to the pole positions of $X(3872)$ in the present framework. For $m_\sigma=400\,\mathrm{MeV}$, $|\Delta E|/\mathrm{B_{con}}$ remains near $0.08$ up to $\delta_a\simeq0.5$ and then rises to about $0.17$ at $\delta_a=1$, showing that the short-range $\delta^{(3)}(\bm r)$ potential generated by the magnetic-dipole term of the vector-meson exchange has a more pronounced effect for this value of $m_\sigma$.

The pole widths in these figures describe the included $D \pi$ and $D \gamma$ constituent-decay mechanisms. While the predicted widths of the $[DD^*]$ pole agree well with the experimental $T_{cc}(3875)^+$ pole, the widths predicted for the $[\bar{D}D^*]$ pole are smaller than the experimental central value (although still within the error bar). This may be because the molecular $[DD^*]$ channel plays a dominant role in the $T_{cc}(3875)^+$ decay, while decays through its compact $cc\bar{u}\bar{d}$ tetraquark component are suppressed relative to the molecular channel. For the $X(3872)$ state, however, the compact components originating from heavy quarkonium $c\bar{c}$ and from the compact tetraquark $c\bar{c}q\bar{q}$ can also contribute to hidden-charm decays, which have not been included so far.


\subsection{Compact-channel mixing and mass shifts} \label{subsec: Feshbach projection}

Since the correlated molecular calculation does not reproduce both pole positions within the parameter range studied. A tension remains in the simultaneous description of $X(3872)$ and $T_{cc}(3875)^+$ as pure molecules. One possible extension is to introduce compact configurations coupled to the molecular channels. The compact channel for $X(3872)$ may originate from charmonium $c\bar{c}$ or from a compact tetraquark $c\bar{c}q\bar{q}$~\cite{Maiani:2004vq,Hogaasen:2005jv,Esposito:2025hlp}, while the compact channel for $T_{cc}(3875)^+$ may originate from a compact $cc\bar{u}\bar{d}$ state~\cite{Chen:2022ros,Kim:2022mpa}. We use a Feshbach projection to illustrate the sign of the mass shift induced by such mixing~\cite{Baru:2010ww,Cincioglu:2016fkm,Kinugawa:2023fbf,Kinugawa:2024crb}. We first reduce the problem to two molecular levels to isolate the mass-shift mechanism.

Representing the molecular sector by two normalized channel configurations, $\ket{1}$ and $\ket{2}$. Their coupled dynamics is understood and can form a bound state first, where
\begin{eqnarray}
H_M|\phi_M\rangle = E_M|\phi_M\rangle , \\
|\phi_M\rangle = c_1|1\rangle+c_2|2\rangle.
\end{eqnarray}
Here $c_i=\langle i|\phi_M\rangle$ is the channel-$i$ component of the molecular eigenstate $|\phi_M\rangle$, and $E_M$ is its unperturbed mass eigenvalue for the pure molecular system. We take $|c_1|^2+|c_2|^2=1 $. The compact state $\ket{c}$ with bare mass $E_c$ couples to the molecular configurations through
\begin{equation}
  g_i=\langle i|V_{\mathrm{mix}}|c\rangle ,
  \qquad i=1,2 \,,
\end{equation}
with $V_{\rm mix}$ the potential coupling the molecular channels to the bare compact channel. For simplicity the matrix elements of couplings $g_i$ are taken to be energy independent. In the basis $\{|1\rangle,|2\rangle,|c\rangle\}$, the total Hamiltonian is
\begin{equation}
  H=
  \begin{pmatrix}
    H_M & \mathbf{g}\\
    \mathbf{g}^\dagger & M_c
  \end{pmatrix}
  =
  \begin{pmatrix}
    H_{11} & H_{12} & g_1\\
    H_{21} & H_{22} & g_2\\
    g_1^* & g_2^* & E_c
  \end{pmatrix}.
 \end{equation}
The effective coupling between the molecular eigenstate and the compact state is the projection of the coupling vector $\mathbf{g}$ onto $|\phi_M\rangle$,
\begin{equation}
  g_{\mathrm{eff}}
  =\langle\phi_M|V_{\mathrm{mix}}|c\rangle
  =\sum_{i=1,2}c_i^*g_i.
\end{equation}
For real $g_i$ and wave functions this reduces to $g_{\mathrm{eff}}=c_1g_1+c_2g_2$. Thus $g_{\mathrm{eff}}$ is the coherent channel projection selected by the molecular eigenstate. To reduce the problem to the molecular and compact sectors, the Feshbach projection operators can be introduced,
\begin{equation}
  P=|\phi_M\rangle\langle\phi_M| ,
  \qquad
  Q=|c\rangle\langle c| ,
  \qquad
  P+Q=1 .
\end{equation}
With
\begin{equation}
  H|\Psi\rangle=E|\Psi\rangle , ~~~   |\psi_P\rangle=P|\Psi\rangle , ~~~   |\psi_Q\rangle=Q|\Psi\rangle ,
\end{equation}
the projected equations are
\begin{eqnarray}
  (E-PHP)|\psi_P\rangle &=& PHQ|\psi_Q\rangle , \\
  (E-QHQ)|\psi_Q\rangle &=& QHP|\psi_P\rangle .
\end{eqnarray}
Eliminating the compact component,
\begin{equation}
  |\psi_Q\rangle=(E-QHQ)^{-1}QHP|\psi_P\rangle ,
\end{equation}
we obtain the effective Schr\"odinger equation
\begin{equation}
  H_{\mathrm{eff}}(E)|\psi_P\rangle=E|\psi_P\rangle ,
\end{equation}
with the Feshbach effective Hamiltonian
\begin{equation}
  H_{\mathrm{eff}}(E)
  =PHP+PHQ\,(E-QHQ)^{-1}QHP .
\end{equation}
For $QHQ=E_cQ$, the eigenvalue condition becomes
\begin{equation}
  E=E_M+\frac{|g_{\mathrm{eff}}|^2}{E-E_c} ,
\end{equation}
with the exact physical eigenvalues are
\begin{equation}
  E_\pm
  =
  \frac{E_M+E_c}{2}
  \pm
  \frac{1}{2}\sqrt{(E_c-E_M)^2+4|g_{\mathrm{eff}}|^2}\, .
  \label{eq:Enegy repultion eigenstates}
\end{equation}

Eq.~\eqref{eq:Enegy repultion eigenstates} exhibits the familiar avoided-crossing level repulsion. The state continuously connected to the molecular level as $g_{\mathrm{eff}}\to0$ is
\begin{equation}
  E_{\mathrm{mol}}=
  \begin{cases}
    E_-, & E_c>E_M \,,\\
    E_+, & E_c<E_M \,.
  \end{cases}
\end{equation}

Therefore the molecular state is pushed away from the compact bare state: for $E_c>E_M$ the molecular-like mass decreases, $E_{\mathrm{mol}}<E_M$, whereas for $E_c<E_M$ it increases, $E_{\mathrm{mol}}>E_M$. In the weak-coupling limit, $|g_{\mathrm{eff}}|\ll|E_M-E_c|$, the shift is
\begin{equation}
  \Delta E_M\equiv E_{\mathrm{mol}}-E_M
  \simeq\frac{|g_{\mathrm{eff}}|^2}{E_M-E_c}.
\end{equation}

Physically, the compact bare state shifts the pure molecular mass, and the resulting eigenstates are superpositions of molecular and compact components. The sign of the shift is controlled by the relative position of the bare compact mass, while its magnitude is set by the projected coupling $|g_{\mathrm{eff}}|$~\cite{Peng:2023lfw}.

For $X(3872)$, a candidate compact channel is the bare $2\,{}^3P_1$ charmonium level. Quark-model estimates typically place it above the $D^0\bar D^{*0}$ threshold; for example, the Godfrey--Isgur model gives approximately $3953\,\mathrm{MeV}$~\cite{Godfrey:1985xj}. Coupling to such a level would lower an already bound molecular level and hence increase the discrepancy in Figs.~\ref{fig:Pole trajectory with only Contact} and \ref{fig:Pole trajectory with OPE} if the fitted molecular interaction were left unchanged.

For $T_{cc}(3875)^+$, if its bare compact mass $E_c<E_M$, the coupling to the compact state would push the physical mass upward. To reproduce the observed $T_{cc}(3875)^+$ pole, $E_M$ would then have to lie deeper than the experimentally observed pole, so that the predicted $[\bar{D}D^*]$ pole would become more deeply bound than in Figs.~\ref{fig:Pole trajectory with only Contact} and \ref{fig:Pole trajectory with OPE}. The inclusion of the bare heavy-quarkonium channel of $X(3872)$, whose mass lies above the molecular mass ($E_c>E_M$), would in addition push the predicted $X(3872)$ mass downward, i.e., further away from the experimental pole.

Conversely, a compact level in $T_{cc}(3875)^+$ above the $D^+D^{*0}$ threshold could supply additional attraction while allowing a weaker underlying molecular interaction. As a numerical illustration of a weaker molecular input, we repeat the iteration with a nominal $[DD^*]$ real energy of $-0.1\,\mathrm{keV}$. Table~\ref{Table: Pole of X with Tcc has 10 keV} gives the results for $\delta_a=1$, $\Lambda=0.8\,\mathrm{GeV}$ and $m_\sigma=400\,\mathrm{MeV}$, which produce the shallowest $[\bar D D^*]$ poles in the scan. The converged $[DD^*]$ pole energies $E_p$ have real parts $-14.9$ and $-16.6\,\mathrm{keV}$ in schemes $A$ and $B$, respectively, rather than the nominal value of $-0.1\,\mathrm{keV}$. The corresponding $D\bar D^*$ poles move closer to threshold but remain about $3\,\mathrm{MeV}$ below it. Scheme $B$ gives $R_X=0.30$, just above the reference band in Eq.~\eqref{eq: RX isospin violation of X3872}.

These results indicate that the $[DD^*]$ system preferentially appears as an unbound state, a virtual state in the coupled $\ket{N_T}$ and $\ket{C_T}$ channels, or a resonance once higher partial waves are included, where a centrifugal barrier can emerge. The $[DD^*]$ potential is then correspondingly shallow, and the linked $[\bar{D}D^*]$ interaction produces a near-threshold $X(3872)$, to which compact-channel mixing may also contribute. Coupling the $DD^*$ channels to a compact $cc\bar{u}\bar{d}$ can then shift the $T_{cc}(3875)^+$ pole toward its experimentally fitted position. If the molecular $DD^*$ state is a bound state with energy $E_M$, the two-level Feshbach analysis places the compact bare level above it, $E_c>E_M$. When $E_M$ lies close to threshold, the compact bare level generally emerges above threshold, and this above-threshold compact bare mass is consistent with the
compact-tetraquark descriptions of $T_{cc}(3875)^+$ obtained in
Refs.~\cite{Kim:2022mpa,Dong:2024upa,Cheng:2020wxa,Faustov:2021hjs,Wang:2022clw}. For a virtual or resonant $DD^*$ pole, however, $E_M$ is defined on an unphysical Riemann sheet and is generally complex, so both the ordering between $E_c$ and $E_M$ and the associated level-repulsion argument must be reexamined on the appropriate sheet.

It is worthy to mention that our results show that the coupled-channel molecular interactions in the $[DD^*]$ and $[\bar{D}D^*]$ systems cannot simultaneously account for the observed $T_{cc}(3875)^+$ and $X(3872)$ as pure bound states: within the pure molecular coupled-channel dynamics, the $T_{cc}(3875)^+$ would preferably be virtual states or resonances, and the compact $cc\bar{u}\bar{d}$ components coupled to the molecular $DD^*$ appear to be essential for forming the bound-like $T_{cc}(3875)^+$ and $X(3872)$ observed experimentally.

\begin{table*}[htbp]
    \renewcommand{\arraystretch}{1.6}
    \setlength{\tabcolsep}{4pt}
    \centering
   \caption{Converged poles obtained with a nominal $[DD^*]$ iteration input of $-0.1\,\mathrm{keV}$, for $\delta_a=1$, $\Lambda=0.8\,\mathrm{GeV}$ and $m_\sigma=400\,\mathrm{MeV}$, together with the predicted $[\bar D D^*]$ poles and $R_X$. Pole energies $E_p$ are measured relative to the $D^0D^{*+}$ and $D^0\bar D^{*0}$ thresholds, respectively. The real parts of the two systems are quoted in different units as indicated.}
   \label{Table: Pole of X with Tcc has 10 keV}
   \begin{tabular}{cccccc}
        \toprule [1.5pt]
        \multirow{2}{*}{Scheme} & \multicolumn{2}{c}{$[DD^*]$} & \multicolumn{3}{c}{$[\bar{D}D^*]$}  \\
         \cmidrule(lr){2-3} \cmidrule(lr){4-6} 
       & $\mathrm{Re}(E_p)$ (keV) & $\mathrm{Im}(E_p)$ (keV) & $\mathrm{Re}(E_p)$ (MeV)  & $\mathrm{Im}(E_p)$ (keV) & $R_X$ \\
     \midrule [1.5pt]
        $A$ & $-14.9$ & $-33.8$ & $-3.01$ & $-8.96$ & $0.17$ \\
        $B$ & $-16.6$ & $-42.8$ & $-3.54$ & $-9.48$ & $0.30$ \\
    \bottomrule  [1.5pt]
    \end{tabular}
\end{table*}

\subsection{Spatial structure in Born--Oppenheimer EFT}

Born-Oppenheimer (BO) effective field theory offers a complementary description of heavy exotic hadrons in terms of heavy-quark motion and light-field configurations~\cite{Brambilla:2017uyf,Bicudo:2019ymo,Soto:2020xpm,Bicudo:2022ihz,Berwein:2024ztx,Braaten:2024tbm}. Its tetraquark channels interpolate between short-distance compact configurations and asymptotic heavy-meson pairs. For the hidden-charm sector, these tetraquark channels can also mix with the separate charmonium channel~\cite{Bicudo:2020qhp,TarrusCastella:2022rxb,TarrusCastella:2024zps,Brambilla:2024imu,Brambilla:2026zfw,Brambilla:2026jqz}. We adopt the BO potential construction of Refs.~\cite{Brambilla:2024imu,Brambilla:2026zfw,Brambilla:2026jqz}. The $X(3872)$ calculation contains a charmonium channel $\ket{X_c}$ and tetraquark channels $\ket{X_S}$ and $\ket{X_D}$ with $S$- and $D$-wave configurations. Spin-isospin averaged heavy-meson masses are used in this description, thus the spin splitting correction is not involved here. For $T_{cc}(3875)^+$, a single tetraquark channel $\ket{T}$ interpolates between the short-distance $cc\bar{u}\bar{d}$ and a meson-meson configuration at large $r$. No charmonium channel occurs in this doubly charmed sector. Here, the coordinate $r$ denotes the separation of the heavy sources, $c\bar c$ for $X(3872)$ and $cc$ for $T_{cc}(3875)^+$.

For $X(3872)$, the mixing potentials between the charmonium and tetraquark
channels near the string-breaking region are constrained by lattice-QCD static-energy calculations~\cite{Brambilla:2024imu,Bulava:2024jpj}. For the tetraquark potentials of $X(3872)$ and $T_{cc}(3875)^+$, the short-distance part is taken from the parametrizations of Ref.~\cite{Alasiri:2024nue}, while
the long-distance tails follow the one-pion-exchange type of
Refs.~\cite{Brambilla:2026zfw,Brambilla:2026jqz}. Once the above potential parameters are fixed to these constructions, the light-field energies $\Lambda_{1^{--}}$ and $\Lambda_{0^+}$ control the locations of the energy levels of the system. $\Lambda_{1^{--}}$ and $\Lambda_{0^+}$ correspond to the adjoint meson mass in the hidden-charm sector and the triplet meson mass in the doubly charmed sector, respectively~\cite{Brambilla:2024imu}.

For $S$-wave two-body scattering at low momentum $k$, the effective-range expansion (ERE) for the phase shift $\delta_0(k)$ reads~\cite{Contessi:2024vae,Peng:2024yzn}
\begin{eqnarray}
    k\,\mathrm{cot}\, \delta_0(k)=-\frac{1}{a_0}+\frac{1}{2}r_0k^2+\mathcal{O}(k^4) \, ,
\end{eqnarray}
where $a_0$ is the scattering length and $r_0$ is the effective range. For the scattering amplitude
\begin{eqnarray}
    f(k)=\frac{1}{k\,\mathrm{cot}\, \delta_0(k)-ik} \, ,
\end{eqnarray}
keeping the leading-order (LO) approximation and taking a stable bound-state pole at $k=i\kappa$ gives
\begin{eqnarray}
    a_0=\frac{1}{\kappa},
\end{eqnarray}
with $\kappa=\sqrt{2\mu_M E_B}$, where $\mu_M$ is the two-body reduced mass and $E_B>0$ is the binding energy. By solving the Schr\"odinger equation within the above BOEFT framework and adjusting the binding energy $E_B$ so as to reproduce the real part of the central values of experimentally extracted scattering lengths of $X(3872)$ and $T_{cc}(3875)^+$~\cite{BESIII:2023hml,LHCb:2021auc},
\begin{eqnarray}
&& X(3872) : a_0^{EX}=  -16.5^{+7.0+5.6}_{-27.6-27.7} \,\,\rm fm \, , \nonumber 
    \\
&& T_{cc}^{+}(3875) :  a_0^{EX}=(-7.16\pm 0.51) +i(1.85\pm0.28)\,\, \rm fm \, , \nonumber
\end{eqnarray}
we obtain the wave functions of the individual channels of $X(3872)$ and $T_{cc}(3875)^+$. The corresponding parameter values are $\Lambda_{1^{--}}=-111.5\,\mathrm{MeV}$ for $X(3872)$ and $\Lambda_{0^+}=-103\,\mathrm{MeV}$ for $T_{cc}^{+}(3875)$.

\begin{figure}[htbp]
    \centering
\includegraphics[width=\linewidth]{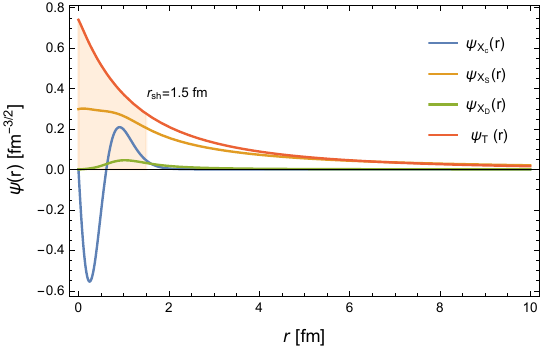}
    \caption{Radial wave functions of $X(3872)$ and $T_{cc}(3875)^+$ obtained within the Born-Oppenheimer effective field theory framework.}
    \label{fig:wave functions from BOEFT}
\end{figure}

As shown in Fig.~\ref{fig:wave functions from BOEFT}, the radial wave functions of $X(3872)$ are $\psi_{X_c}(r)$, $\psi_{X_S}(r)$, and $\psi_{X_D}(r)$ for the channels $\ket{X_c}$, $\ket{X_S}$, and $\ket{X_{D}}$, respectively. For $T_{cc}(3875)^+$, $\psi_T(r)$ is the wave function of the single tetraquark channel, which receives contributions from both the compact-tetraquark and the meson-meson configurations. The wave functions extend to large distances, even beyond $r>10 \, \rm fm$, a clear signature of the molecule-dominated structure of $X(3872)$ and $T_{cc}(3875)^+$. Moreover, the small magnitude of $\psi_{X_D}(r)$ indicates that the $S$-wave tetraquark channel is much more important than the $D$-wave one in $X(3872)$.

Since the heavy-quarkonium channel $\ket{X_c}$ of $X(3872)$ is represented explicitly among the BOEFT degrees of freedom, its compact contribution is quantified by the probability obtained by integrating the corresponding wave function over the entire range of $r$,
\begin{eqnarray}
   P_{X_c} =\int r^2dr\,|\psi_{X_c}(r)|^2=2.66\% \,.
\end{eqnarray}
By contrast, the tetraquark channels of $X(3872)$ and $T_{cc}(3875)^+$ are described by wave functions that interpolate between a compact configuration at short distances and a molecular configuration at large distances, both living in the same coordinate-space Hilbert space in $r$. The compact and molecular probabilities cannot therefore be defined unambiguously by integrating over the entire range of $r$: the molecular component contributes at short distances, and the compact component leaves an imprint at long distances through these tetraquark wave functions.

We therefore introduce a separation scale $r_{\rm sh}$ of the order of the typical hadronic size, $r_{\rm sh}=0.5$--$1.5\,\mathrm{fm}$, and define the short- and long-distance tetraquark weights as
\begin{eqnarray}
    P_{X_{S;C}} &=& \int_0^{r_{\rm{sh}}} r^2dr\,|\psi_{X_S}(r)|^2 \, , \\
    P_{X_{S;M}} &=& \int_{r_{\rm{sh}}}^{\infty} r^2dr\,|\psi_{X_S}(r)|^2 \,,    \\
   P_{X_{D;C}} &=& \int_0^{r_{\rm{sh}}} r^2dr\,|\psi_{X_D}(r)|^2 \, , \\
    P_{X_{D;M}} &=& \int_{r_{\rm{sh}}}^{\infty} r^2dr\,|\psi_{X_D}(r)|^2 \, .
\end{eqnarray}
The subscripts $C$ and $M$ label the short and long spatial regions in this prescription, respectively. We then use these short- and long-distance tetraquark weights for a phenomenological estimation of the compact-tetraquark and molecular probabilities. The combined charmonium and short-distance tetraquark weight for $X(3872)$ is
\begin{eqnarray}
P_{X_C}=P_{X_c}+P_{X_{S;C}}+P_{X_{D;C}} \, .
\end{eqnarray}

Similarly, for the single-channel wave function $\psi_T(r)$ of $T_{cc}(3875)^+$, the short- and long distance tetraquark weights are
\begin{eqnarray}
    P_{T_{C}} &=& \int_0^{r_{\rm{sh}}} r^2dr\,|\psi_{T}(r)|^2 \, , \\
    P_{T_{M}} &=& \int_{r_{\rm{sh}}}^{\infty} r^2dr\,|\psi_{T}(r)|^2 \, .
\end{eqnarray}

The values of these wights $P_{X_{C}}$ and $P_{T_{C}}$, estimated with the separation scale $r_{\rm sh}$, which we used as estimation of compact probabilities, are listed in Table~\ref{Table: Probabilities of rsh}. As $r_{\rm sh}$ increases from $0.5 \, \rm fm$ to $1.5 \, \rm fm$, $P_{T_C}$ grows from $1.34\%$ to $14.4\%$, a more dramatic variation than that of the compact probability of $X(3872)$, which ranges from $3.02\%$ to $9.66\%$; this points to nontrivial short-range dynamics in $T_{cc}(3875)^+$. In any case, the sizable values of $P_{X_{C}}$ and $P_{T_{C}}$ reflect the importance of the compact components within $X(3872)$ and $T_{cc}(3875)^+$, consistent with the pole-trajectory results for the pure $[\bar{D}D^*]$ interpretation of $X(3872)$ with the $[DD^*]$ input for $T_{cc}(3875)^+$ discussed in Sec.~\ref{subsec: Pole trajectories} and \ref{subsec: Feshbach projection}.

\begin{table}[htbp]
    \renewcommand{\arraystretch}{1.6}
    \setlength{\tabcolsep}{4pt}
    \centering
   \caption{Compact probabilities $P_{X_{C}}$ and $P_{T_{C}}$ and the corresponding molecular compositeness for $X(3872)$ and $T_{cc}(3875)^+$, for $r_{\rm sh}=0.5$, $1$, and $1.5\, \mathrm{fm}$.}
   \begin{tabular}{lrrrr}
        \toprule [1.5pt]
        \multirow{2}{*}{$r_{\rm sh} \,\rm (fm)$} & \multicolumn{2}{c}{$X(3872)$} & \multicolumn{2}{c}{$T_{cc}(3875)^+$}  \\
        \cmidrule(lr){2-3} \cmidrule(lr){4-5}
       & $P_{X_C}$  & $1-P_{X_C}$ & $P_{T_C}$   & $1-P_{T_C}$  \\
     \midrule [1.5pt]
        $0.5$ & $3.02\%$ & $96.98\%$ & $1.34\%$ & $98.66\%$ \\
        $1$ & $5.31\%$ & $94.69\%$ & $6.58\%$ & $93.42\%$  \\
        $1.5$ & $9.66\%$ & $90.34\%$ & $14.41\%$ & $85.59\%$ \\
    \bottomrule [1.5pt]
    \end{tabular}     \label{Table: Probabilities of rsh}
\end{table}

\section{Summary and discussion} \label{sec5: summary and discussions}

We have investigated short-range dynamics in the coupled $[\bar D D^*]$ and $[DD^*]$ systems associated with $X(3872)$ and $T_{cc}(3875)^+$. The interaction consists of contact terms and an explicit one-pion-exchange potential with its local contribution subtracted. Energy-dependent $D^*$ self-energies and the time-ordered OPE kernel incorporate the associated three-body cuts~\cite{Baru:2011rs,Du:2021zzh,Zhang:2024fxy}. Scalar- and vector-meson exchange saturation, together with a common renormalization-group scaling prescription, relates the contact couplings in the two systems~\cite{Peng:2020xrf,Peng:2021hkr,Yang:2022ezl,Peng:2023lfw}. The real part of the $T_{cc}(3875)^+$ pole~\cite{LHCb:2021auc} fixes the overall interaction strength, allowing the $X(3872)$ pole to be predicted.

The local contribution generated by the magnetic-dipole coupling of vector mesons plays a central role in this correlation. Its exclusion from saturation relies on separating its short-distance scale from the range of the exchanged meson~\cite{Peng:2020xrf}. For $\rho$ and $\omega$ exchanges, however, the vector-meson masses and the characteristic hadronic form-factor scale are comparable. The finite-range exchange and the smeared local contribution therefore belong to the same short-distance region, without a strong scale hierarchy that would justify neglecting the latter. Retaining the full local term also preserves the cancellation of the momentum-independent contributions in the low-momentum expansion of the complete magnetic-exchange potential. These considerations motivate including this local contribution in the short-range matching and examining its effect through $\delta_a$. Increasing $\delta_a$ reduces the $X(3872)$ attraction relative to that in the $T_{cc}(3875)^+$ system after calibration, and full retention, $\delta_a=1$, produces the shallowest $X(3872)$ poles in the parameter scan. In scheme $B$, the closest solutions still lie about $5$--$12~\mathrm{MeV}$ below the observed $X(3872)$ pole. Retaining the local term therefore substantially reduces the overbinding but does not fully resolve the discrepancy between the predicted and observed pole positions. This result extends the tension identified in the earlier single-channel saturation study~\cite{Peng:2023lfw} to a coupled-channel calculation that includes explicit pion exchange and three-body effects.

Pion exchange has a moderate effect on the binding energy and a more pronounced effect on the isospin-violating coupling ratio. At $\Lambda=1~\mathrm{GeV}$, the relative binding-energy difference between schemes $A$ and $B$ is approximately $8$--$9\%$ over much of the scan and reaches $17\%$ for $m_\sigma=400~\mathrm{MeV}$ at $\delta_a=1$. Scheme $B$ brings the residue-based ratio $R_X$ into the phenomenological range $0.26 \pm 0.03$~\cite{Dias:2024zfh} for part of the scan, whereas scheme $A$ remains below it. The neutral- and charged-channel residues are therefore more sensitive to explicit pion exchange than the pole energy~\cite{Hidalgo-Duque:2012rqv,Zhang:2024fxy,Ji:2025hjw}. In the $T_{cc}(3875)^+$ system, the pion exchange increases the predicted half-width from approximately $17$ to $28~\mathrm{keV}$, close to the experimental central value of $24~\mathrm{keV}$~\cite{LHCb:2021auc,Du:2021zzh,Qiu:2023uno} within uncertainties.

The Feshbach analysis shows how compact-channel mixing can modify the mass correlation~\cite{Baru:2010ww,Cincioglu:2016fkm,Kinugawa:2023fbf,Kinugawa:2024crb}.
When a compact level lies above a molecular level, the molecular-like eigenvalue is shifted downward. Consequently, coupling the present overbound $X(3872)$ solution to an above-threshold bare $2\,{}^3P_1$ charmonium level~\cite{Godfrey:1985xj} would further increase the discrepancy if the molecular interaction were kept fixed. 
A simultaneous description of the two states therefore requires refitting the molecular interaction together with the compact-channel couplings. In the $T_{cc}(3875)^+$ sector, a compact $cc\bar{u}\bar{d}$ channel can provide additional attraction and allows for a weaker molecular interaction within the same saturation scheme. A unified treatment of the compact states and the molecular configurations would determine the resulting pole structure.

With the energy scales constrained by the real parts of the experimental scattering lengths~\cite{BESIII:2023hml,LHCb:2021auc}, the Born--Oppenheimer effective field theory wave functions provide a complementary account of the spatial structure~\cite{Brambilla:2024imu,Brambilla:2026zfw,Brambilla:2026jqz}. The charmonium-channel weight in $X(3872)$ is $2.66\%$. Adding the short-distance portions of the tetraquark channels gives $P_{X_C}=3.02$--$9.66\%$ for separation radii $r_{\rm sh}=0.5$--$1.5~\mathrm{fm}$. For $T_{cc}(3875)^+$, the corresponding short-distance weight is $P_{T_C}=1.34$--$14.41\%$. Both states are dominated by extended meson-pair configurations, with nonzero short-distance contributions. Together with the Feshbach analysis, these results support a description in which molecular channels coexists with compact dynamics that can influence the near-threshold poles.

The present interaction contains leading contact interaction and OPE potential. A natural extension is to include derivative contact operators and higher-order pion dynamics, particularly two-pion exchange (TPE). EFT and lattice QCD studies have highlighted the role of TPE in the interaction~\cite{Lu:2021gsb,Wang:2018atz,Wang:2022jop,Zhai:2023ejo,Geng:2026ppg}, especially in the $DD^*$ system~\cite{Lyu:2023xro,Xu:2025xrl}. Its low-momentum analytic contributions can be matched onto contact operators at the corresponding order, while the remaining nonanalytic part supplies an explicit finite-range interaction~\cite{Chacko:2024cax}. Consistent matching would clarify the relation between phenomenological $\sigma$ exchange and TPE, avoid double counting, and quantify their effect on our spectrum analysis.

Our results favor retaining the local magnetic-dipole contribution in the vector-meson exchange saturation scheme and show that its treatment substantially affects the correlation between the $X(3872)$ and $T_{cc}(3875)^+$ poles. The residual discrepancy between the predicted $[\bar{D} D^*]$ poles and the observed $X(3872)$ pole, together with the compact-channel mixing analysis and the BOEFT wave functions, motivates a joint description of molecular and compact dynamics; in particular, $T_{cc}(3875)^+$ should be coupled to a compact $cc\bar{u}\bar{d}$ state.

\begin{acknowledgments}

This work is partly supported by the National Key R\&D Program of China under Grant No. 2023YFA1606703, by the Postdoctoral Fellowship Program of CPSF under Grant No. GZC20241765, by the National Natural Science Foundation of China under Grant Nos. 12435007, W2543006, 12361141819, and 12575094. F.-Z.P. acknowledges support from the Alexander von Humboldt Foundation.

\end{acknowledgments}

\bibliographystyle{apsrev4-2}
\bibliography{reference}

@article{Lu:2021gsb,
    author = "Lu, Jun-Xu and Wang, Chun-Xuan and Xiao, Yang and Geng, Li-Sheng and Meng, Jie and Ring, Peter",
    title = "{Accurate Relativistic Chiral Nucleon-Nucleon Interaction up to Next-to-Next-to-Leading Order}",
    eprint = "2111.07766",
    archivePrefix = "arXiv",
    primaryClass = "nucl-th",
    doi = "10.1103/PhysRevLett.128.142002",
    journal = "Phys. Rev. Lett.",
    volume = "128",
    number = "14",
    pages = "142002",
    year = "2022"
}

@article{Kinugawa:2023fbf,
    author = "Kinugawa, Tomona and Hyodo, Tetsuo",
    title = "{Compositeness of Tcc and X(3872) by considering decay and coupled-channels effects}",
    eprint = "2303.07038",
    archivePrefix = "arXiv",
    primaryClass = "hep-ph",
    doi = "10.1103/PhysRevC.109.045205",
    journal = "Phys. Rev. C",
    volume = "109",
    number = "4",
    pages = "045205",
    year = "2024"
}

@article{Maiani:2004vq,
    author = "Maiani, L. and Piccinini, F. and Polosa, A. D. and Riquer, V.",
    title = "{Diquark-antidiquarks with hidden or open charm and the nature of X(3872)}",
    eprint = "hep-ph/0412098",
    archivePrefix = "arXiv",
    reportNumber = "ROMA1-1396-2004, FNT-T-2004-20, BA-TH-502-04, CERN-PH-TH-2004-239",
    doi = "10.1103/PhysRevD.71.014028",
    journal = "Phys. Rev. D",
    volume = "71",
    pages = "014028",
    year = "2005"
}

@article{Hogaasen:2005jv,
    author = "Hogaasen, H. and Richard, J. M. and Sorba, P.",
    title = "{A Chromomagnetic mechanism for the X(3872) resonance}",
    eprint = "hep-ph/0511039",
    archivePrefix = "arXiv",
    reportNumber = "LAPTH-1221-05, LPSC-05-112",
    doi = "10.1103/PhysRevD.73.054013",
    journal = "Phys. Rev. D",
    volume = "73",
    pages = "054013",
    year = "2006"
}

@article{Kim:2022mpa,
    author = "Kim, Yonghee and Oka, Makoto and Suzuki, Kei",
    title = "{Doubly heavy tetraquarks in a chiral-diquark picture}",
    eprint = "2202.06520",
    archivePrefix = "arXiv",
    primaryClass = "hep-ph",
    doi = "10.1103/PhysRevD.105.074021",
    journal = "Phys. Rev. D",
    volume = "105",
    number = "7",
    pages = "074021",
    year = "2022"
}

@article{Chen:2022ros,
    author = "Chen, Xiaoyun and Wang, Fu-Lai and Tan, Yue and Yang, Youchang",
    title = "{Double-heavy tetraquarks with strangeness in the chiral quark model*}",
    eprint = "2206.10917",
    archivePrefix = "arXiv",
    primaryClass = "hep-ph",
    doi = "10.1088/1674-1137/ac9de9",
    journal = "Chin. Phys. C",
    volume = "47",
    number = "2",
    pages = "023102",
    year = "2023"
}

@article{Dong:2024upa,
    author = "Dong, Wen-Chao and Wang, Zhi-Gang",
    title = "{Hunting for the prospective Tcc family based on the diquark-antidiquark configuration}",
    eprint = "2407.19383",
    archivePrefix = "arXiv",
    primaryClass = "hep-ph",
    doi = "10.1016/j.nuclphysb.2025.116828",
    journal = "Nucl. Phys. B",
    volume = "1012",
    pages = "116828",
    year = "2025"
}

@article{Cheng:2020wxa,
    author = "Cheng, Jian-Bo and Li, Shi-Yuan and Liu, Yan-Rui and Si, Zong-Guo and Yao, Tao",
    title = "{Double-heavy tetraquark states with heavy diquark-antiquark symmetry}",
    eprint = "2008.00737",
    archivePrefix = "arXiv",
    primaryClass = "hep-ph",
    doi = "10.1088/1674-1137/abde2f",
    journal = "Chin. Phys. C",
    volume = "45",
    number = "4",
    pages = "043102",
    year = "2021"
}

@article{Faustov:2021hjs,
    author = "Faustov, Rudolf N. and Galkin, Vladimir O. and Savchenko, Elena M.",
    title = "{Heavy tetraquarks in the relativistic quark model}",
    eprint = "2103.01763",
    archivePrefix = "arXiv",
    primaryClass = "hep-ph",
    doi = "10.3390/universe7040094",
    journal = "Universe",
    volume = "7",
    number = "4",
    pages = "94",
    year = "2021"
}

@article{Wang:2022clw,
    author = "Wang, Jin-Bao and Li, Gang and An, Chun-Sheng and Deng, Cheng-Rong and Xie, Ju-Jun",
    title = "{The low-lying hidden- and double-charm tetraquark states in a constituent quark model with instanton-induced interaction}",
    eprint = "2204.13320",
    archivePrefix = "arXiv",
    primaryClass = "hep-ph",
    doi = "10.1140/epjc/s10052-022-10673-7",
    journal = "Eur. Phys. J. C",
    volume = "82",
    number = "8",
    pages = "721",
    year = "2022"
}

@article{Chacko:2024cax,
    author = "Chacko, J. T. and Baru, V. and Hanhart, C. and Krug, S. L.",
    title = "{Two-pion exchange for coupled-channel scattering of two heavy mesons}",
    eprint = "2411.13303",
    archivePrefix = "arXiv",
    doi = "10.1103/PhysRevD.111.034042",
    journal = "Phys. Rev. D",
    volume = "111",
    number = "3",
    pages = "034042",
    year = "2025"
}

@article{Zhai:2023ejo,
    author = "Zhai, Qing-Yu and Liu, Ming-Zhu and Lu, Jun-Xu and Geng, Li-Sheng",
    title = "{Long-range S-wave DD* interaction in covariant chiral effective field theory}",
    eprint = "2311.06569",
    archivePrefix = "arXiv",
    primaryClass = "hep-ph",
    doi = "10.1103/PhysRevD.109.034015",
    journal = "Phys. Rev. D",
    volume = "109",
    number = "3",
    pages = "034015",
    year = "2024"
}

@article{Geng:2026ppg,
    author = "Geng, Li-Sheng and Lu, Jun-Xu and Zhai, Qing-Yu and Liu, Zhi-Wei and Shen, Shi-Hang",
    title = "{Recent Developments in and Applications of the Relativistic Chiral Nuclear Force}",
    eprint = "2603.01819",
    archivePrefix = "arXiv",
    primaryClass = "nucl-th",
    doi = "10.3390/particles9020038",
    journal = "Particles",
    volume = "9",
    number = "2",
    pages = "38",
    year = "2026"
}

@article{Valderrama:2012jv,
    author = "Valderrama, M. Pavon",
    title = "{Power Counting and Perturbative One Pion Exchange in Heavy Meson Molecules}",
    eprint = "1204.2400",
    archivePrefix = "arXiv",
    primaryClass = "hep-ph",
    doi = "10.1103/PhysRevD.85.114037",
    journal = "Phys. Rev. D",
    volume = "85",
    pages = "114037",
    year = "2012"
}

@article{Wang:2018atz,
    author = "Wang, Bo and Liu, Zhan-Wei and Liu, Xiang",
    title = "{$\bar{B}^{(\ast)} \bar{B}^{(\ast)}$ interactions in chiral effective field theory}",
    eprint = "1812.04457",
    archivePrefix = "arXiv",
    primaryClass = "hep-ph",
    doi = "10.1103/PhysRevD.99.036007",
    journal = "Phys. Rev. D",
    volume = "99",
    number = "3",
    pages = "036007",
    year = "2019"
}

@article{Xu:2025xrl,
    author = "Xu, Hao and Ren, Li-xiang",
    title = "{Investigating the two-pion exchange of the double charm DD* chiral interactions and Tcc}",
    eprint = "2509.11234",
    archivePrefix = "arXiv",
    primaryClass = "hep-ph",
    doi = "10.1103/gryg-xtt4",
    journal = "Phys. Rev. D",
    volume = "112",
    number = "11",
    pages = "114003",
    year = "2025"
}

@article{Wang:2022jop,
    author = "Wang, Bo and Meng, Lu",
    title = "{Revisiting the DD* chiral interactions with the local momentum-space regularization up to the third order and the nature of Tcc+}",
    eprint = "2212.08447",
    archivePrefix = "arXiv",
    primaryClass = "hep-ph",
    doi = "10.1103/PhysRevD.107.094002",
    journal = "Phys. Rev. D",
    volume = "107",
    number = "9",
    pages = "094002",
    year = "2023"
}

@article{Lyu:2023xro,
    author = "Lyu, Yan and Aoki, Sinya and Doi, Takumi and Hatsuda, Tetsuo and Ikeda, Yoichi and Meng, Jie",
    title = "{Doubly Charmed Tetraquark Tcc+ from Lattice QCD near Physical Point}",
    eprint = "2302.04505",
    archivePrefix = "arXiv",
    primaryClass = "hep-lat",
    reportNumber = "RIKEN-iTHEMS-Report-23, YITP-23-14",
    doi = "10.1103/PhysRevLett.131.161901",
    journal = "Phys. Rev. Lett.",
    volume = "131",
    number = "16",
    pages = "161901",
    year = "2023"
}

@article{BESIII:2023hml,
    author = "Ablikim, Medina and others",
    collaboration = "BESIII",
    title = "{Coupled-Channel Analysis of the {\ensuremath{\chi}}c1(3872) Line Shape with BESIII Data}",
    eprint = "2309.01502",
    archivePrefix = "arXiv",
    primaryClass = "hep-ex",
    doi = "10.1103/PhysRevLett.132.151903",
    journal = "Phys. Rev. Lett.",
    volume = "132",
    number = "15",
    pages = "151903",
    year = "2024"
}

@article{Peng:2024yzn,
    author = "Peng, Fang-Zheng and Geng, Li-Sheng and Xie, Ju-Jun",
    title = "{Constraining the hidden-charm pentaquark predictions and discriminating the Pc(4440) and Pc(4457) spins through the effective range expansion}",
    eprint = "2410.21891",
    archivePrefix = "arXiv",
    primaryClass = "hep-ph",
    doi = "10.1103/PhysRevD.111.054029",
    journal = "Phys. Rev. D",
    volume = "111",
    number = "5",
    pages = "054029",
    year = "2025"
}

@article{Bulava:2024jpj,
    author = "Bulava, John and Knechtli, Francesco and Koch, Vanessa and Morningstar, Colin and Peardon, Michael",
    title = "{The quark-mass dependence of the potential energy between static colour sources in the QCD vacuum with light and strange quarks}",
    eprint = "2403.00754",
    archivePrefix = "arXiv",
    primaryClass = "hep-lat",
    reportNumber = "WUB/24-00",
    doi = "10.1016/j.physletb.2024.138754",
    journal = "Phys. Lett. B",
    volume = "854",
    pages = "138754",
    year = "2024"
}

@article{Alasiri:2024nue,
    author = "Alasiri, Fareed and Braaten, Eric and Mohapatra, Abhishek",
    title = "{Born-Oppenheimer potentials for SU(3) gauge theory}",
    eprint = "2406.05123",
    archivePrefix = "arXiv",
    primaryClass = "hep-ph",
    reportNumber = "TUM-EFT 188/24",
    doi = "10.1103/PhysRevD.110.054029",
    journal = "Phys. Rev. D",
    volume = "110",
    number = "5",
    pages = "054029",
    year = "2024"
}

@article{Brambilla:2026jqz,
    author = "Brambilla, Nora and Bruschini, Roberto and Mohapatra, Abhishek and Peng, Fang-Zheng and Scirpa, Tommaso",
    title = "{Quarkoniumlike states above open-flavor thresholds in Born-Oppenheimer EFT}",
    journal = "",
    eprint = "2608.04105",
    archivePrefix = "arXiv",
    primaryClass = "hep-ph",
    month = "8",
    year = "2026"
}

@article{Bicudo:2020qhp,
    author = "Bicudo, Pedro and Cardoso, Nuno and Mueller, Lasse and Wagner, Marc",
    title = "{Computation of the quarkonium and meson-meson composition of the $\Upsilon(nS)$  states and of the new $\Upsilon(10753)$ Belle resonance from lattice QCD static potentials}",
    eprint = "2008.05605",
    archivePrefix = "arXiv",
    primaryClass = "hep-lat",
    doi = "10.1103/PhysRevD.103.074507",
    journal = "Phys. Rev. D",
    volume = "103",
    number = "7",
    pages = "074507",
    year = "2021"
}

@article{Bicudo:2022ihz,
    author = "Bicudo, Pedro and Cardoso, Nuno and Mueller, Lasse and Wagner, Marc",
    title = "{Study of I=0 bottomonium bound states and resonances in S, P, D, and F waves with lattice QCD static-static-light-light potentials}",
    eprint = "2205.11475",
    archivePrefix = "arXiv",
    primaryClass = "hep-lat",
    doi = "10.1103/PhysRevD.107.094515",
    journal = "Phys. Rev. D",
    volume = "107",
    number = "9",
    pages = "094515",
    year = "2023"
}

@article{Bicudo:2019ymo,
    author = "Bicudo, Pedro and Cardoso, Marco and Cardoso, Nuno and Wagner, Marc",
    title = "{Bottomonium resonances with $I = 0$ from lattice QCD correlation functions with static and light quarks}",
    eprint = "1910.04827",
    archivePrefix = "arXiv",
    primaryClass = "hep-lat",
    doi = "10.1103/PhysRevD.101.034503",
    journal = "Phys. Rev. D",
    volume = "101",
    number = "3",
    pages = "034503",
    year = "2020"
}

@article{Brambilla:2026zfw,
    author = "Brambilla, Nora and Mohapatra, Abhishek and Scirpa, Tommaso and Vairo, Antonio",
    title = "{Open-flavor threshold effects on quarkonium spectrum in the BOEFT}",
    eprint = "2604.12603",
    archivePrefix = "arXiv",
    primaryClass = "hep-ph",
    reportNumber = "TUM-EFT 199/25",
    journal = "",
    month = "4",
    year = "2026"
}

@article{Brambilla:2024imu,
    author = "Brambilla, Nora and Mohapatra, Abhishek and Scirpa, Tommaso and Vairo, Antonio",
    title = "{Nature of $\chi_{c1}(3872)$ and $T_{cc}^+(3875)$}",
    eprint = "2411.14306",
    archivePrefix = "arXiv",
    primaryClass = "hep-ph",
    reportNumber = "TUM-EFT 193/24",
    doi = "10.1103/pdy7-hvg7",
    journal = "Phys. Rev. Lett.",
    volume = "135",
    number = "13",
    pages = "131902",
    year = "2025"
}

@article{Braaten:2024tbm,
    author = "Braaten, Eric and Bruschini, Roberto",
    title = "{Exotic hidden-heavy hadrons and where to find them}",
    eprint = "2409.08002",
    archivePrefix = "arXiv",
    primaryClass = "hep-ph",
    doi = "10.1016/j.physletb.2025.139386",
    journal = "Phys. Lett. B",
    volume = "863",
    pages = "139386",
    year = "2025"
}

@article{TarrusCastella:2024zps,
    author = "Tarr\'us Castell\`a, Jaume",
    title = "{Effect of continuum states on the double-heavy hadron spectra}",
    eprint = "2401.13393",
    archivePrefix = "arXiv",
    primaryClass = "hep-ph",
    doi = "10.1007/JHEP06(2024)107",
    journal = "JHEP",
    volume = "06",
    pages = "107",
    year = "2024"
}

@article{TarrusCastella:2022rxb,
    author = "Tarr\'us Castell\`a, Jaume",
    title = "{Heavy meson thresholds in Born-Oppenheimer effective field theory}",
    eprint = "2207.09365",
    archivePrefix = "arXiv",
    primaryClass = "hep-ph",
    doi = "10.1103/PhysRevD.106.094020",
    journal = "Phys. Rev. D",
    volume = "106",
    number = "9",
    pages = "094020",
    year = "2022"
}

@article{Berwein:2024ztx,
    author = "Berwein, Matthias and Brambilla, Nora and Mohapatra, Abhishek and Vairo, Antonio",
    title = "{Hybrids, tetraquarks, pentaquarks, doubly heavy baryons, and quarkonia in Born-Oppenheimer effective theory}",
    eprint = "2408.04719",
    archivePrefix = "arXiv",
    primaryClass = "hep-ph",
    reportNumber = "TUM-EFT 185/23",
    doi = "10.1103/PhysRevD.110.094040",
    journal = "Phys. Rev. D",
    volume = "110",
    number = "9",
    pages = "094040",
    year = "2024"
}

@article{Soto:2020xpm,
    author = "Soto, Joan and Tarr{\'u}s Castell{\`a}, Jaume",
    title = "{Nonrelativistic effective field theory for heavy exotic hadrons}",
    eprint = "2005.00552",
    archivePrefix = "arXiv",
    primaryClass = "hep-ph",
    doi = "10.1103/PhysRevD.102.014012",
    journal = "Phys. Rev. D",
    volume = "102",
    number = "1",
    pages = "014012",
    year = "2020",
    note = "[Erratum: Phys.Rev.D 110, 099901 (2024)]"
}

@article{Brambilla:2017uyf,
    author = "Brambilla, Nora and Krein, Gast{\~a}o and Tarr{\'u}s Castell{\`a}, Jaume and Vairo, Antonio",
    title = "{Born-Oppenheimer approximation in an effective field theory language}",
    eprint = "1707.09647",
    archivePrefix = "arXiv",
    primaryClass = "hep-ph",
    reportNumber = "TUM-EFT-69-15",
    doi = "10.1103/PhysRevD.97.016016",
    journal = "Phys. Rev. D",
    volume = "97",
    number = "1",
    pages = "016016",
    year = "2018"
}

@article{Cincioglu:2016fkm,
    author = "Cincioglu, E. and Nieves, J. and Ozpineci, A. and Yilmazer, A. U.",
    title = "{Quarkonium Contribution to Meson Molecules}",
    eprint = "1606.03239",
    archivePrefix = "arXiv",
    primaryClass = "hep-ph",
    doi = "10.1140/epjc/s10052-016-4413-1",
    journal = "Eur. Phys. J. C",
    volume = "76",
    number = "10",
    pages = "576",
    year = "2016"
}

@article{Kinugawa:2024crb,
    author = "Kinugawa, Tomona and Hyodo, Tetsuo",
    title = "{Compositeness of hadrons, nuclei, and atomic systems}",
    eprint = "2411.12285",
    archivePrefix = "arXiv",
    primaryClass = "hep-ph",
    doi = "10.1140/epja/s10050-025-01548-y",
    journal = "Eur. Phys. J. A",
    volume = "61",
    number = "7",
    pages = "154",
    year = "2025"
}

@article{Baru:2010ww,
    author = "Baru, V. and Hanhart, C. and Kalashnikova, Yu. S. and Kudryavtsev, A. E. and Nefediev, A. V.",
    title = "{Interplay of quark and meson degrees of freedom in a near-threshold resonance}",
    eprint = "1001.0369",
    archivePrefix = "arXiv",
    primaryClass = "hep-ph",
    reportNumber = "FZJ-IKP-TH-2009-39",
    doi = "10.1140/epja/i2010-10929-7",
    journal = "Eur. Phys. J. A",
    volume = "44",
    pages = "93--103",
    year = "2010"
}

@article{Baru:2011rs,
    author = "Baru, V. and Filin, A. A. and Hanhart, C. and Kalashnikova, Yu. S. and Kudryavtsev, A. E. and Nefediev, A. V.",
    title = "{Three-body $D\bar{D}\pi$ dynamics for the X(3872)}",
    eprint = "1108.5644",
    archivePrefix = "arXiv",
    primaryClass = "hep-ph",
    doi = "10.1103/PhysRevD.84.074029",
    journal = "Phys. Rev. D",
    volume = "84",
    pages = "074029",
    year = "2011"
}

@article{Doring:2009yv,
    author = "Doring, M. and Hanhart, C. and Huang, F. and Krewald, S. and Meissner, U. -G.",
    title = "{Analytic properties of the scattering amplitude and resonances parameters in a meson exchange model}",
    eprint = "0903.4337",
    archivePrefix = "arXiv",
    primaryClass = "nucl-th",
    reportNumber = "FZJ-IKP-TH-2009-11, HISKP-TH-09-14",
    doi = "10.1016/j.nuclphysa.2009.08.010",
    journal = "Nucl. Phys. A",
    volume = "829",
    pages = "170--209",
    year = "2009"
}

@article{Dong:2024fjk,
    author = "Dong, Xiang-Kun and Ji, Teng and Guo, Feng-Kun and Mei{\ss}ner, Ulf-G. and Zou, Bing-Song",
    title = "{Hints of the JPC = 0{\ensuremath{-}}{\ensuremath{-}} and 1{\ensuremath{-}}{\ensuremath{-}}K{\textasteriskcentered}K{\textasciimacron}1(1270) molecules in the J/{\ensuremath{\psi}} {\textrightarrow} {\ensuremath{\phi}}{\ensuremath{\eta}}{\ensuremath{\eta}}' decay}",
    eprint = "2402.02903",
    archivePrefix = "arXiv",
    primaryClass = "hep-ph",
    doi = "10.1016/j.physletb.2024.138646",
    journal = "Phys. Lett. B",
    volume = "853",
    pages = "138646",
    year = "2024"
}

@article{ParticleDataGroup:2026mpi,
    author = "Takahashi, F. and others",
    collaboration = "Particle Data Group",
    title = "{Review of Particle Physics}",
    doi = "10.1142/s0217751x26300115",
    journal = "Int. J. Mod. Phys. A",
    volume = "41",
    number = "22",
    pages = "2630011",
    year = "2026"
}

@article{Valderrama:2016koj,
    author = "Valderrama, Manuel Pavon",
    title = "{Power Counting and Wilsonian Renormalization in Nuclear Effective Field Theory}",
    eprint = "1604.01332",
    archivePrefix = "arXiv",
    primaryClass = "nucl-th",
    doi = "10.1142/S021830131641007X",
    journal = "Int. J. Mod. Phys. E",
    volume = "25",
    number = "05",
    pages = "1641007",
    year = "2016"
}

@article{PavonValderrama:2014zeq,
    author = "Pav{\'o}n Valderrama, M. and Phillips, Daniel R.",
    title = "{Power Counting of Contact-Range Currents in Effective Field Theory}",
    eprint = "1407.0437",
    archivePrefix = "arXiv",
    primaryClass = "nucl-th",
    doi = "10.1103/PhysRevLett.114.082502",
    journal = "Phys. Rev. Lett.",
    volume = "114",
    number = "8",
    pages = "082502",
    year = "2015"
}

@article{Yan:2026zvw,
    author = "Yan, Mao-Jun",
    title = "{Decoding the near-threshold \(X_{0,1}(4140)\) and \(X_{1}(4685)\) states via OZI-suppressed coupled-channel scattering}",
    eprint = "2602.17125",
    archivePrefix = "arXiv",
    primaryClass = "hep-ph",
    doi = "10.1103/1p1p-ysfd",
    journal = "Phys. Rev. D",
    volume = "113",
    number = "9",
    pages = "094012",
    year = "2026"
}

@article{Wang:2022vjm,
    author = "Wang, Hao-Nan and Wang, Qian and Xie, Ju-Jun",
    title = "{Theoretical study on the contributions of {\ensuremath{\omega}} meson to the X(3872){\textrightarrow}J/{\ensuremath{\psi}}{\ensuremath{\pi}}+{\ensuremath{\pi}}- and J/{\ensuremath{\psi}}{\ensuremath{\pi}}+{\ensuremath{\pi}}-{\ensuremath{\pi}}0 decays}",
    eprint = "2206.14456",
    archivePrefix = "arXiv",
    primaryClass = "hep-ph",
    doi = "10.1103/PhysRevD.106.056022",
    journal = "Phys. Rev. D",
    volume = "106",
    number = "5",
    pages = "056022",
    year = "2022"
}

@article{Ge:2026moy,
    author = "Ge, Duo-Lun and Liu, Zhi-Wei and Geng, Li-Sheng",
    title = "{DD* correlation functions in deciphering the nature of Tcc(3875)+}",
    eprint = "2603.24980",
    archivePrefix = "arXiv",
    primaryClass = "hep-ph",
    doi = "10.1103/xgks-ft1w",
    journal = "Phys. Rev. D",
    volume = "114",
    number = "1",
    pages = "014054",
    year = "2026"
}

@article{Wang:2022xga,
    author = "Wang, Hao-Nan and Geng, Li-Sheng and Wang, Qian and Xie, Ju-Jun",
    title = "{Molecular Nature of X(3872) in B $^{0} \to K ^{0}$ X(3872) and B $^{+}$ K $^{+}$ X(3872) Decays}",
    eprint = "2211.14994",
    archivePrefix = "arXiv",
    primaryClass = "hep-ph",
    doi = "10.1088/0256-307X/40/2/021301",
    journal = "Chin. Phys. Lett.",
    volume = "40",
    number = "2",
    pages = "021301",
    year = "2023"
}

@article{Peng:2020hql,
    author = "Peng, Fang-Zheng and Yan, Mao-Jun and S{\'a}nchez S{\'a}nchez, Mario and Valderrama, Manuel Pavon",
    title = "{The $P_{cs}(4459)$ pentaquark from a combined effective field theory and phenomenological perspective}",
    eprint = "2011.01915",
    archivePrefix = "arXiv",
    primaryClass = "hep-ph",
    doi = "10.1140/epjc/s10052-021-09416-x",
    journal = "Eur. Phys. J. C",
    volume = "81",
    number = "7",
    pages = "666",
    year = "2021"
}

@article{Qiu:2023uno,
    author = "Qiu, Lin and Gong, Chang and Zhao, Qiang",
    title = "{Coupled-channel description of charmed heavy hadronic molecules within the meson-exchange model and its implication}",
    eprint = "2311.10067",
    archivePrefix = "arXiv",
    primaryClass = "hep-ph",
    doi = "10.1103/PhysRevD.109.076016",
    journal = "Phys. Rev. D",
    volume = "109",
    number = "7",
    pages = "076016",
    year = "2024"
}

@article{Peng:2023lfw,
    author = "Peng, Fang-Zheng and Yan, Mao-Jun and Pavon Valderrama, Manuel",
    title = "{Heavy- and light-flavor symmetry partners of the Tcc+(3875), the X(3872), and the X(3960) from light-meson exchange saturation}",
    eprint = "2304.13515",
    archivePrefix = "arXiv",
    primaryClass = "hep-ph",
    doi = "10.1103/PhysRevD.108.114001",
    journal = "Phys. Rev. D",
    volume = "108",
    number = "11",
    pages = "114001",
    year = "2023"
}

@article{Yan:2023ttx,
    author = "Yan, Mao-Jun and Peng, Fang-Zheng and Pavon Valderrama, Manuel",
    title = "{Molecular charmed baryons and pentaquarks from light-meson exchange saturation}",
    eprint = "2304.14855",
    archivePrefix = "arXiv",
    primaryClass = "hep-ph",
    doi = "10.1103/PhysRevD.109.014023",
    journal = "Phys. Rev. D",
    volume = "109",
    number = "1",
    pages = "014023",
    year = "2024"
}

@article{Yang:2022ezl,
    author = "Yang, Zi-Ying and Peng, Fang-Zheng and Yan, Mao-Jun and S{\'a}nchez S{\'a}nchez, Mario and Pavon Valderrama, Manuel",
    title = "{Molecular P{\ensuremath{\psi}} pentaquarks from light-meson exchange saturation}",
    eprint = "2211.08211",
    archivePrefix = "arXiv",
    primaryClass = "hep-ph",
    doi = "10.1103/PhysRevD.111.014012",
    journal = "Phys. Rev. D",
    volume = "111",
    number = "1",
    pages = "014012",
    year = "2025"
}

@article{Peng:2021hkr,
    author = "Peng, Fang-Zheng and S{\'a}nchez S{\'a}nchez, Mario and Yan, Mao-Jun and Pavon Valderrama, Manuel",
    title = "{Heavy-hadron molecular spectrum from light-meson exchange saturation}",
    eprint = "2101.07213",
    archivePrefix = "arXiv",
    primaryClass = "hep-ph",
    doi = "10.1103/PhysRevD.105.034028",
    journal = "Phys. Rev. D",
    volume = "105",
    number = "3",
    pages = "034028",
    year = "2022"
}

@article{Yan:2021wdl,
    author = "Yan, Mao-Jun and Valderrama, Manuel Pavon",
    title = "{Subleading contributions to the decay width of the Tcc+ tetraquark}",
    eprint = "2108.04785",
    archivePrefix = "arXiv",
    primaryClass = "hep-ph",
    doi = "10.1103/PhysRevD.105.014007",
    journal = "Phys. Rev. D",
    volume = "105",
    number = "1",
    pages = "014007",
    year = "2022"
}

@article{Du:2021zzh,
    author = "Du, Meng-Lin and Baru, Vadim and Dong, Xiang-Kun and Filin, Arseniy and Guo, Feng-Kun and Hanhart, Christoph and Nefediev, Alexey and Nieves, Juan and Wang, Qian",
    title = "{Coupled-channel approach to Tcc+ including three-body effects}",
    eprint = "2110.13765",
    archivePrefix = "arXiv",
    primaryClass = "hep-ph",
    doi = "10.1103/PhysRevD.105.014024",
    journal = "Phys. Rev. D",
    volume = "105",
    number = "1",
    pages = "014024",
    year = "2022"
}

@article{Albaladejo:2021vln,
    author = "Albaladejo, M.",
    title = "{Tcc+ coupled channel analysis and predictions}",
    eprint = "2110.02944",
    archivePrefix = "arXiv",
    primaryClass = "hep-ph",
    doi = "10.1016/j.physletb.2022.137052",
    journal = "Phys. Lett. B",
    volume = "829",
    pages = "137052",
    year = "2022"
}

@article{Feijoo:2021ppq,
    author = "Feijoo, A. and Liang, W. H. and Oset, Eulogio",
    title = "{D0D0{\ensuremath{\pi}}+ mass distribution in the production of the Tcc exotic state}",
    eprint = "2108.02730",
    archivePrefix = "arXiv",
    primaryClass = "hep-ph",
    doi = "10.1103/PhysRevD.104.114015",
    journal = "Phys. Rev. D",
    volume = "104",
    number = "11",
    pages = "114015",
    year = "2021"
}

@article{LHCb:2021auc,
    author = "Aaij, Roel and others",
    collaboration = "LHCb",
    title = "{Study of the doubly charmed tetraquark $T_{cc}^{+}$}",
    eprint = "2109.01056",
    archivePrefix = "arXiv",
    primaryClass = "hep-ex",
    reportNumber = "CERN-EP-2021-169, LHCb-PAPER-2021-032",
    doi = "10.1038/s41467-022-30206-w",
    journal = "Nature Commun.",
    volume = "13",
    number = "1",
    pages = "3351",
    year = "2022"
}

@article{LHCb:2021vvq,
    author = "Aaij, Roel and others",
    collaboration = "LHCb",
    title = "{Observation of an exotic narrow doubly charmed tetraquark}",
    eprint = "2109.01038",
    archivePrefix = "arXiv",
    primaryClass = "hep-ex",
    reportNumber = "CERN-EP-2021-165, LHCb-PAPER-2021-031",
    doi = "10.1038/s41567-022-01614-y",
    journal = "Nature Phys.",
    volume = "18",
    number = "7",
    pages = "751--754",
    year = "2022"
}

@article{Zhang:2024fxy,
    author = "Zhang, Zhen-Hua and Ji, Teng and Dong, Xiang-Kun and Guo, Feng-Kun and Hanhart, Christoph and Mei{\ss}ner, Ulf-G. and Rusetsky, Akaki",
    title = "{Predicting isovector charmonium-like states from X(3872) properties}",
    eprint = "2404.11215",
    archivePrefix = "arXiv",
    primaryClass = "hep-ph",
    doi = "10.1007/JHEP08(2024)130",
    journal = "JHEP",
    volume = "08",
    pages = "130",
    year = "2024"
}

@article{Ji:2025hjw,
    author = "Ji, Teng and Dong, Xiang-Kun and Guo, Feng-Kun and Hanhart, Christoph and Mei{\ss}ner, Ulf-G.",
    title = "{Precise determination of the properties of X(3872) and of its isovector partner Wc1}",
    eprint = "2502.04458",
    archivePrefix = "arXiv",
    primaryClass = "hep-ph",
    doi = "10.1016/j.scib.2026.07.071",
    journal = "Sci. Bull.",
    volume = "71",
    pages = "4416--4421",
    year = "2026"
}

@article{Dong:2026qbr,
    author = "Dong, Xiang-Kun and Guo, Feng-Kun and Hanhart, Christoph and Ji, Teng and Mei{\ss}ner, Ulf-G.",
    title = "{Amplified Isospin Breaking from Coupled-Channel Dynamics Near Threshold}",
    journal = "",
    eprint = "2608.16266",
    archivePrefix = "arXiv",
    primaryClass = "hep-ph",
    month = "8",
    year = "2026"
}

@article{Meng:2021kmi,
    author = "Meng, Lu and Wang, Guang-Juan and Wang, Bo and Zhu, Shi-Lin",
    title = "{Revisit the isospin violating decays of X(3872)}",
    eprint = "2109.01333",
    archivePrefix = "arXiv",
    primaryClass = "hep-ph",
    doi = "10.1103/PhysRevD.104.094003",
    journal = "Phys. Rev. D",
    volume = "104",
    number = "9",
    pages = "094003",
    year = "2021"
}

@article{Gamermann:2009uq,
    author = "Gamermann, D. and Nieves, J. and Oset, E. and Ruiz Arriola, E.",
    title = "{Couplings in coupled channels versus wave functions: application to the X(3872) resonance}",
    eprint = "0911.4407",
    archivePrefix = "arXiv",
    primaryClass = "hep-ph",
    doi = "10.1103/PhysRevD.81.014029",
    journal = "Phys. Rev. D",
    volume = "81",
    pages = "014029",
    year = "2010"
}

@article{Li:2012cs,
    author = "Li, Ning and Zhu, Shi-Lin",
    title = "{Isospin breaking, Coupled-channel effects and Diagnosis of X(3872)}",
    eprint = "1207.3954",
    archivePrefix = "arXiv",
    primaryClass = "hep-ph",
    doi = "10.1103/PhysRevD.86.074022",
    journal = "Phys. Rev. D",
    volume = "86",
    pages = "074022",
    year = "2012"
}

@article{Gamermann:2009fv,
    author = "Gamermann, Daniel and Oset, Eulogio",
    title = "{Isospin breaking effects in the X(3872) resonance}",
    eprint = "0905.0402",
    archivePrefix = "arXiv",
    primaryClass = "hep-ph",
    doi = "10.1103/PhysRevD.80.014003",
    journal = "Phys. Rev. D",
    volume = "80",
    pages = "014003",
    year = "2009"
}

@article{Hidalgo-Duque:2012rqv,
    author = "Hidalgo-Duque, C. and Nieves, J. and Valderrama, M. Pavon",
    title = "{Light flavor and heavy quark spin symmetry in heavy meson molecules}",
    eprint = "1210.5431",
    archivePrefix = "arXiv",
    primaryClass = "hep-ph",
    doi = "10.1103/PhysRevD.87.076006",
    journal = "Phys. Rev. D",
    volume = "87",
    number = "7",
    pages = "076006",
    year = "2013"
}

@article{Albaladejo:2015dsa,
    author = "Albaladejo, M. and Guo, F. -K. and Hidalgo-Duque, C. and Nieves, J. and Valderrama, M. Pavon",
    title = "{Decay widths of the spin-2 partners of the X(3872)}",
    eprint = "1504.00861",
    archivePrefix = "arXiv",
    primaryClass = "hep-ph",
    doi = "10.1140/epjc/s10052-015-3753-6",
    journal = "Eur. Phys. J. C",
    volume = "75",
    number = "11",
    pages = "547",
    year = "2015"
}

@article{Contessi:2020jqa,
    author = "Contessi, Lorenzo and Kirscher, Johannes and Pavon Valderrama, Manuel",
    title = "{Triple-$X$ and beyond: hadronic systems of three and more $X(3872)$}",
    eprint = "2008.12268",
    archivePrefix = "arXiv",
    primaryClass = "hep-ph",
    doi = "10.1103/PhysRevD.103.056001",
    journal = "Phys. Rev. D",
    volume = "103",
    number = "5",
    pages = "056001",
    year = "2021"
}

@article{Nieves:2011zz,
    author = "Nieves, J. and Valderrama, M. Pavon",
    title = "{Deriving the existence of $B\bar{B}^*$ bound states from the X(3872) and Heavy Quark Symmetry}",
    eprint = "1106.0600",
    archivePrefix = "arXiv",
    primaryClass = "hep-ph",
    doi = "10.1103/PhysRevD.84.056015",
    journal = "Phys. Rev. D",
    volume = "84",
    pages = "056015",
    year = "2011"
}

@article{Nieves:2012tt,
    author = "Nieves, J. and Valderrama, M. Pavon",
    title = "{The Heavy Quark Spin Symmetry Partners of the X(3872)}",
    eprint = "1204.2790",
    archivePrefix = "arXiv",
    primaryClass = "hep-ph",
    doi = "10.1103/PhysRevD.86.056004",
    journal = "Phys. Rev. D",
    volume = "86",
    pages = "056004",
    year = "2012"
}

@article{Guo:2014hqa,
    author = "Guo, F. K. and Hidalgo-Duque, C. and Nieves, J. and Ozpineci, Altug and Valderrama, M. P.",
    title = "{Detecting the long-distance structure of the $X$(3872)}",
    eprint = "1404.1776",
    archivePrefix = "arXiv",
    primaryClass = "hep-ph",
    doi = "10.1140/epjc/s10052-014-2885-4",
    journal = "Eur. Phys. J. C",
    volume = "74",
    number = "5",
    pages = "2885",
    year = "2014"
}

@article{Kalashnikova:2018vkv,
    author = "Kalashnikova, Yu. S. and Nefediev, A. V.",
    title = "{X(3872) in the molecular model}",
    eprint = "1811.01324",
    archivePrefix = "arXiv",
    primaryClass = "hep-ph",
    doi = "10.3367/UFNe.2018.08.038411",
    journal = "Phys. Usp.",
    volume = "62",
    number = "6",
    pages = "568--595",
    year = "2019"
}

@article{Esposito:2025hlp,
    author = "Esposito, Angelo and Glioti, Alfredo and Germani, Davide and Polosa, Antonio D.",
    title = "{A short review on the compositeness of the X(3872)}",
    eprint = "2502.02505",
    archivePrefix = "arXiv",
    primaryClass = "hep-ph",
    doi = "10.1007/s40766-025-00066-3",
    journal = "Riv. Nuovo Cim.",
    volume = "48",
    number = "2",
    pages = "95--155",
    year = "2025"
}

@article{Guo:2013zbw,
    author = "Guo, Feng-Kun and Hanhart, Christoph and Mei{\ss}ner, Ulf-G. and Wang, Qian and Zhao, Qiang",
    title = "{Production of the X(3872) in charmonia radiative decays}",
    eprint = "1306.3096",
    archivePrefix = "arXiv",
    primaryClass = "hep-ph",
    doi = "10.1016/j.physletb.2013.06.053",
    journal = "Phys. Lett. B",
    volume = "725",
    pages = "127--133",
    year = "2013"
}

@article{Lee:2009hy,
    author = "Lee, Ian Woo and Faessler, Amand and Gutsche, Thomas and Lyubovitskij, Valery E.",
    title = "{X(3872) as a molecular DD* state in a potential model}",
    eprint = "0910.1009",
    archivePrefix = "arXiv",
    primaryClass = "hep-ph",
    doi = "10.1103/PhysRevD.80.094005",
    journal = "Phys. Rev. D",
    volume = "80",
    pages = "094005",
    year = "2009"
}

@article{Braaten:2004fk,
    author = "Braaten, Eric and Kusunoki, Masaoki and Nussinov, Shmuel",
    title = "{Production of the X(3870) in B meson decay by the coalescence of charm mesons}",
    eprint = "hep-ph/0404161",
    archivePrefix = "arXiv",
    doi = "10.1103/PhysRevLett.93.162001",
    journal = "Phys. Rev. Lett.",
    volume = "93",
    pages = "162001",
    year = "2004"
}

@article{LHCb:2013kgk,
    author = "Aaij, R and others",
    collaboration = "LHCb",
    title = "{Determination of the X(3872) meson quantum numbers}",
    eprint = "1302.6269",
    archivePrefix = "arXiv",
    primaryClass = "hep-ex",
    reportNumber = "LHCB-PAPER-2013-001, CERN-PH-EP-2013-017",
    doi = "10.1103/PhysRevLett.110.222001",
    journal = "Phys. Rev. Lett.",
    volume = "110",
    pages = "222001",
    year = "2013"
}

@article{BESIII:2022bse,
    author = "Ablikim, M. and others",
    collaboration = "BESIII",
    title = "{Observation of a New X(3872) Production Process e+e-{\textrightarrow}{\ensuremath{\omega}}X(3872)}",
    eprint = "2212.07291",
    archivePrefix = "arXiv",
    primaryClass = "hep-ex",
    doi = "10.1103/PhysRevLett.130.151904",
    journal = "Phys. Rev. Lett.",
    volume = "130",
    number = "15",
    pages = "151904",
    year = "2023"
}

@article{LHCb:2020xds,
    author = "Aaij, R. and others",
    collaboration = "LHCb",
    title = "{Study of the lineshape of the $\chi_{c1}(3872)$ state}",
    eprint = "2005.13419",
    archivePrefix = "arXiv",
    primaryClass = "hep-ex",
    reportNumber = "CERN-EP-2020-086, LHCb-PAPER-2020-008",
    doi = "10.1103/PhysRevD.102.092005",
    journal = "Phys. Rev. D",
    volume = "102",
    number = "9",
    pages = "092005",
    year = "2020"
}

@article{CMS:2021znk,
    author = "Sirunyan, Albert M. and others",
    collaboration = "CMS",
    title = "{Evidence for X(3872) in Pb-Pb Collisions and Studies of its Prompt Production at $\sqrt {s_{NN}}$=5.02{\,}{\,}TeV}",
    eprint = "2102.13048",
    archivePrefix = "arXiv",
    primaryClass = "hep-ex",
    reportNumber = "CMS-HIN-19-005, CERN-EP-2021-023",
    doi = "10.1103/PhysRevLett.128.032001",
    journal = "Phys. Rev. Lett.",
    volume = "128",
    number = "3",
    pages = "032001",
    year = "2022"
}

@article{ATLAS:2016kwu,
    author = "Aaboud, Morad and others",
    collaboration = "ATLAS",
    title = "{Measurements of $\psi(2S)$ and $X(3872) \to J/\psi\pi^+\pi^-$ production in $pp$ collisions at $\sqrt{s} = 8$ TeV with the ATLAS detector}",
    eprint = "1610.09303",
    archivePrefix = "arXiv",
    primaryClass = "hep-ex",
    reportNumber = "CERN-EP-2016-193",
    doi = "10.1007/JHEP01(2017)117",
    journal = "JHEP",
    volume = "01",
    pages = "117",
    year = "2017"
}

@article{BESIII:2019esk,
    author = "Ablikim, M. and others",
    collaboration = "BESIII",
    title = "{Observation of the decay $X(3872) \to \pi^0 \chi_{c1}(1P)$}",
    eprint = "1901.03992",
    archivePrefix = "arXiv",
    primaryClass = "hep-ex",
    doi = "10.1103/PhysRevLett.122.202001",
    journal = "Phys. Rev. Lett.",
    volume = "122",
    number = "20",
    pages = "202001",
    year = "2019"
}

@article{BESIII:2013fnz,
    author = "Ablikim, M. and others",
    collaboration = "BESIII",
    title = "{Observation of $e^+e^− → γX$(3872) at BESIII}",
    eprint = "1310.4101",
    archivePrefix = "arXiv",
    primaryClass = "hep-ex",
    doi = "10.1103/PhysRevLett.112.092001",
    journal = "Phys. Rev. Lett.",
    volume = "112",
    number = "9",
    pages = "092001",
    year = "2014"
}

@article{BaBar:2010wfc,
    author = "del Amo Sanchez, P. and others",
    collaboration = "BaBar",
    title = "{Evidence for the decay X(3872) ---{\ensuremath{>}} J/ psi omega}",
    eprint = "1005.5190",
    archivePrefix = "arXiv",
    primaryClass = "hep-ex",
    reportNumber = "SLAC-PUB-14143, BABAR-PUB-10-009",
    doi = "10.1103/PhysRevD.82.011101",
    journal = "Phys. Rev. D",
    volume = "82",
    pages = "011101",
    year = "2010"
}

@article{Belle:2006olv,
    author = "Gokhroo, G. and others",
    collaboration = "Belle",
    title = "{Observation of a Near-threshold D0 anti-D0 pi0 Enhancement in B ---{\ensuremath{>}} D0 anti-D0 pi0 K Decay}",
    eprint = "hep-ex/0606055",
    archivePrefix = "arXiv",
    reportNumber = "KEK-PREPRINT-2006-16, BELLE-PREPRINT-2006-19",
    doi = "10.1103/PhysRevLett.97.162002",
    journal = "Phys. Rev. Lett.",
    volume = "97",
    pages = "162002",
    year = "2006"
}

@article{D0:2004zmu,
    author = "Abazov, V. M. and others",
    collaboration = "D0",
    title = "{Observation and properties of the $X(3872)$ decaying to $J/\psi \pi^+ \pi^-$ in $p\bar{p}$ collisions at $\sqrt{s} = 1.96$ TeV}",
    eprint = "hep-ex/0405004",
    archivePrefix = "arXiv",
    reportNumber = "FERMILAB-PUB-04-061-E",
    doi = "10.1103/PhysRevLett.93.162002",
    journal = "Phys. Rev. Lett.",
    volume = "93",
    pages = "162002",
    year = "2004"
}

@article{Belle:2003nnu,
    author = "Choi, S. K. and others",
    collaboration = "Belle",
    title = "{Observation of a narrow charmonium-like state in exclusive $B^\pm \to K^\pm \pi^+ \pi^- J/\psi$ decays}",
    eprint = "hep-ex/0309032",
    archivePrefix = "arXiv",
    doi = "10.1103/PhysRevLett.91.262001",
    journal = "Phys. Rev. Lett.",
    volume = "91",
    pages = "262001",
    year = "2003"
}

@article{CalleCordon:2009pit,
    author = "Calle Cordon, A. and Ruiz Arriola, E.",
    title = "{Renormalization vs Strong Form Factors for One Boson Exchange Potentials}",
    eprint = "0905.4933",
    archivePrefix = "arXiv",
    primaryClass = "nucl-th",
    doi = "10.1103/PhysRevC.81.044002",
    journal = "Phys. Rev. C",
    volume = "81",
    pages = "044002",
    year = "2010"
}

@article{Meng:2017fwb,
    author = "Meng, Lu and Li, Ning and Zhu, Shi-Lin",
    title = "{Deuteron-like states composed of two doubly charmed baryons}",
    eprint = "1704.01009",
    archivePrefix = "arXiv",
    primaryClass = "hep-ph",
    doi = "10.1103/PhysRevD.95.114019",
    journal = "Phys. Rev. D",
    volume = "95",
    number = "11",
    pages = "114019",
    year = "2017"
}

@article{Machleidt:1987hj,
    author = "Machleidt, R. and Holinde, K. and Elster, C.",
    title = "{The Bonn Meson Exchange Model for the Nucleon Nucleon Interaction}",
    doi = "10.1016/S0370-1573(87)80002-9",
    journal = "Phys. Rept.",
    volume = "149",
    pages = "1--89",
    year = "1987"
}

@article{Ling:2021asz,
    author = "Ling, Xi-Zhe and Liu, Ming-Zhu and Geng, Li-Sheng",
    title = "{Masses and strong decays of open charm hexaquark states $\Sigma _{c}^{(*)}{\Sigma }_{c}^{(*)}$}",
    eprint = "2110.13792",
    archivePrefix = "arXiv",
    primaryClass = "hep-ph",
    doi = "10.1140/epjc/s10052-021-09867-2",
    journal = "Eur. Phys. J. C",
    volume = "81",
    number = "12",
    pages = "1090",
    year = "2021"
}

@article{Liu:2019zvb,
    author = "Liu, Ming-Zhu and Wu, Tian-Wei and S{\'a}nchez S{\'a}nchez, Mario and Valderrama, Manuel Pavon and Geng, Li-Sheng and Xie, Ju-Jun",
    title = "{Spin-parities of the $P_c(4440)$ and $P_c(4457)$ in the one-boson-exchange model}",
    eprint = "1907.06093",
    archivePrefix = "arXiv",
    primaryClass = "hep-ph",
    doi = "10.1103/PhysRevD.103.054004",
    journal = "Phys. Rev. D",
    volume = "103",
    number = "5",
    pages = "054004",
    year = "2021"
}

@article{Thomas:2008ja,
    author = "Thomas, C. E. and Close, F. E.",
    title = "{Is X(3872) a molecule?}",
    eprint = "0805.3653",
    archivePrefix = "arXiv",
    primaryClass = "hep-ph",
    reportNumber = "OUTP-08-08P",
    doi = "10.1103/PhysRevD.78.034007",
    journal = "Phys. Rev. D",
    volume = "78",
    pages = "034007",
    year = "2008"
}

@article{Wang:2024ukc,
    author = "Wang, Jun-Zhang and Lin, Zi-Yang and Wang, Bo and Meng, Lu and Zhu, Shi-Lin",
    title = "{Double pole structures of X1(2900) as the P-wave D{\textasciimacron}*K* resonances}",
    eprint = "2408.08965",
    archivePrefix = "arXiv",
    primaryClass = "hep-ph",
    doi = "10.1103/PhysRevD.110.114003",
    journal = "Phys. Rev. D",
    volume = "110",
    number = "11",
    pages = "114003",
    year = "2024"
}

@article{Chen:2021tip,
    author = "Chen, Rui",
    title = "{Strong decays of the newly $P_{cs}(4459)$ as a strange hidden-charm $\Xi _c{\bar{D}}^*$ molecule}",
    eprint = "2101.10614",
    archivePrefix = "arXiv",
    primaryClass = "hep-ph",
    doi = "10.1140/epjc/s10052-021-08904-4",
    journal = "Eur. Phys. J. C",
    volume = "81",
    number = "2",
    pages = "122",
    year = "2021"
}

@article{Wang:2020dya,
    author = "Wang, Fu-Lai and Liu, Xiang",
    title = "{Exotic double-charm molecular states with hidden or open strangeness and around $4.5\sim 4.7$ GeV}",
    eprint = "2008.13484",
    archivePrefix = "arXiv",
    primaryClass = "hep-ph",
    doi = "10.1103/PhysRevD.102.094006",
    journal = "Phys. Rev. D",
    volume = "102",
    number = "9",
    pages = "094006",
    year = "2020"
}

@article{Braaten:2020iye,
    author = "Braaten, Eric and He, Li-Ping and Ingles, Kevin and Jiang, Jun",
    title = "{Charm-meson triangle singularity in ${e^+e^-}$ annihilation into ${ D^{*0} \bar{D}^0 + \gamma }$}",
    eprint = "2004.12841",
    archivePrefix = "arXiv",
    primaryClass = "hep-ph",
    doi = "10.1103/PhysRevD.101.096020",
    journal = "Phys. Rev. D",
    volume = "101",
    number = "9",
    pages = "096020",
    year = "2020"
}

@article{Rosner:2013sha,
    author = "Rosner, Jonathan L.",
    title = "{Hadronic and radiative $D^*$ widths}",
    eprint = "1307.2550",
    archivePrefix = "arXiv",
    primaryClass = "hep-ph",
    reportNumber = "EFI-13-13",
    doi = "10.1103/PhysRevD.88.034034",
    journal = "Phys. Rev. D",
    volume = "88",
    number = "3",
    pages = "034034",
    year = "2013"
}

@article{Guo:2019qcn,
    author = "Guo, Feng-Kun",
    title = "{Novel Method for Precisely Measuring the $X(3872)$ Mass}",
    eprint = "1902.11221",
    archivePrefix = "arXiv",
    primaryClass = "hep-ph",
    doi = "10.1103/PhysRevLett.122.202002",
    journal = "Phys. Rev. Lett.",
    volume = "122",
    number = "20",
    pages = "202002",
    year = "2019"
}

@article{Liu:2009qhy,
    author = "Liu, Xiang and Luo, Zhi-Gang and Liu, Yan-Rui and Zhu, Shi-Lin",
    title = "{X(3872) and Other Possible Heavy Molecular States}",
    eprint = "0808.0073",
    archivePrefix = "arXiv",
    primaryClass = "hep-ph",
    doi = "10.1140/epjc/s10052-009-1020-4",
    journal = "Eur. Phys. J. C",
    volume = "61",
    pages = "411--428",
    year = "2009"
}

@article{Wang:2022mxy,
    author = "Wang, Fu-Lai and Liu, Xiang",
    title = "{Emergence of molecular-type characteristic spectrum of hidden-charm pentaquark with strangeness embodied in the P{\ensuremath{\psi}}s{\ensuremath{\Lambda}}(4338) and Pcs(4459)}",
    eprint = "2207.10493",
    archivePrefix = "arXiv",
    primaryClass = "hep-ph",
    doi = "10.1016/j.physletb.2022.137583",
    journal = "Phys. Lett. B",
    volume = "835",
    pages = "137583",
    year = "2022"
}

@article{Zou:2025sae,
    author = "Zou, Bing-Song",
    title = "{Establishing the vector-meson-exchange dominance for the short range interactions of light quarks}",
    eprint = "2502.04414",
    archivePrefix = "arXiv",
    primaryClass = "hep-ph",
    doi = "10.1007/s11433-024-2561-2",
    journal = "Sci. China Phys. Mech. Astron.",
    volume = "68",
    pages = "232031",
    year = "2025"
}

@article{Yalikun:2025ssz,
    author = "Yalikun, Nijiati and Dong, Xiang-Kun and Mei{\ss}ner, Ulf-G.",
    title = "{Role of the short-range dynamics in the formation of D(*)D{\textasciimacron}(*) and B(*)B{\textasciimacron}(*) hadronic molecules}",
    eprint = "2503.01322",
    archivePrefix = "arXiv",
    primaryClass = "hep-ph",
    doi = "10.1103/PhysRevD.111.094036",
    journal = "Phys. Rev. D",
    volume = "111",
    number = "9",
    pages = "094036",
    year = "2025"
}

@article{Wang:2025ioe,
    author = "Wang, Ziye and Yalikun, Nijiati and Reyimuaji, Yakefu",
    title = "{Role of the short-range dynamics in a simultaneous interpretation of Pcs pentaquarks as {\ensuremath{\Xi}}c(',*)D{\textasciimacron}(*) molecules}",
    eprint = "2505.02601",
    archivePrefix = "arXiv",
    primaryClass = "hep-ph",
    doi = "10.1103/p4m3-ml1k",
    journal = "Phys. Rev. C",
    volume = "112",
    number = "4",
    pages = "045205",
    year = "2025"
}

@article{PavonValderrama:2019ixb,
    author = "Pavon Valderrama, Manuel",
    title = "{Heavy hadron molecules in effective field theory: the emergence of exotic nuclear landscapes}",
    eprint = "1906.06491",
    archivePrefix = "arXiv",
    primaryClass = "hep-ph",
    doi = "10.1140/epja/s10050-020-00099-8",
    journal = "Eur. Phys. J. A",
    volume = "56",
    number = "4",
    pages = "109",
    year = "2020"
}

@article{Liu:2019stu,
    author = "Liu, Ming-Zhu and Wu, Tian-Wei and Pavon Valderrama, Manuel and Xie, Ju-Jun and Geng, Li-Sheng",
    title = "{Heavy-quark spin and flavor symmetry partners of the X(3872) revisited: What can we learn from the one boson exchange model?}",
    eprint = "1902.03044",
    archivePrefix = "arXiv",
    primaryClass = "hep-ph",
    doi = "10.1103/PhysRevD.99.094018",
    journal = "Phys. Rev. D",
    volume = "99",
    number = "9",
    pages = "094018",
    year = "2019"
}

@article{PavonValderrama:2019nbk,
    author = "Pavon Valderrama, Manuel",
    title = "{One pion exchange and the quantum numbers of the P$_c$(4440) and P$_c$(4457) pentaquarks}",
    eprint = "1907.05294",
    archivePrefix = "arXiv",
    primaryClass = "hep-ph",
    doi = "10.1103/PhysRevD.100.094028",
    journal = "Phys. Rev. D",
    volume = "100",
    number = "9",
    pages = "094028",
    year = "2019"
}

@article{Ding:2008gr,
    author = "Ding, Gui-Jun",
    title = "{Are Y(4260) and Z+(2) are D(1) D or D(0) D* Hadronic Molecules?}",
    eprint = "0809.4818",
    archivePrefix = "arXiv",
    primaryClass = "hep-ph",
    doi = "10.1103/PhysRevD.79.014001",
    journal = "Phys. Rev. D",
    volume = "79",
    pages = "014001",
    year = "2009"
}

@article{Chen:2016spr,
    author = "Chen, Hua-Xing and Chen, Wei and Liu, Xiang and Liu, Yan-Rui and Zhu, Shi-Lin",
    title = "{A review of the open charm and open bottom systems}",
    eprint = "1609.08928",
    archivePrefix = "arXiv",
    primaryClass = "hep-ph",
    doi = "10.1088/1361-6633/aa6420",
    journal = "Rept. Prog. Phys.",
    volume = "80",
    number = "7",
    pages = "076201",
    year = "2017"
}

@article{Liu:2019zoy,
    author = "Liu, Yan-Rui and Chen, Hua-Xing and Chen, Wei and Liu, Xiang and Zhu, Shi-Lin",
    title = "{Pentaquark and Tetraquark states}",
    eprint = "1903.11976",
    archivePrefix = "arXiv",
    primaryClass = "hep-ph",
    doi = "10.1016/j.ppnp.2019.04.003",
    journal = "Prog. Part. Nucl. Phys.",
    volume = "107",
    pages = "237--320",
    year = "2019"
}

@article{Chen:2022asf,
    author = "Chen, Hua-Xing and Chen, Wei and Liu, Xiang and Liu, Yan-Rui and Zhu, Shi-Lin",
    title = "{An updated review of the new hadron states}",
    eprint = "2204.02649",
    archivePrefix = "arXiv",
    primaryClass = "hep-ph",
    doi = "10.1088/1361-6633/aca3b6",
    journal = "Rept. Prog. Phys.",
    volume = "86",
    number = "2",
    pages = "026201",
    year = "2023"
}

@article{Meng:2022ozq,
    author = "Meng, Lu and Wang, Bo and Wang, Guang-Juan and Zhu, Shi-Lin",
    title = "{Chiral perturbation theory for heavy hadrons and chiral effective field theory for heavy hadronic molecules}",
    eprint = "2204.08716",
    archivePrefix = "arXiv",
    primaryClass = "hep-ph",
    doi = "10.1016/j.physrep.2023.04.003",
    journal = "Phys. Rept.",
    volume = "1019",
    pages = "1--149",
    year = "2023"
}

@article{Reinert:2017usi,
    author = "Reinert, P. and Krebs, H. and Epelbaum, E.",
    title = "{Semilocal momentum-space regularized chiral two-nucleon potentials up to fifth order}",
    eprint = "1711.08821",
    archivePrefix = "arXiv",
    primaryClass = "nucl-th",
    doi = "10.1140/epja/i2018-12516-4",
    journal = "Eur. Phys. J. A",
    volume = "54",
    number = "5",
    pages = "86",
    year = "2018"
}

@article{Peng:2020xrf,
    author = "Peng, Fang-Zheng and Liu, Ming-Zhu and S{\'a}nchez S{\'a}nchez, Mario and Pavon Valderrama, Manuel",
    title = "{Heavy-hadron molecules from light-meson-exchange saturation}",
    eprint = "2004.05658",
    archivePrefix = "arXiv",
    primaryClass = "hep-ph",
    doi = "10.1103/PhysRevD.102.114020",
    journal = "Phys. Rev. D",
    volume = "102",
    pages = "114020",
    year = "2020"
}

@article{Dias:2024zfh,
    author = "Dias, Jorgivan Morais and Ji, Teng and Dong, Xiang-Kun and Guo, Feng-Kun and Hanhart, Christoph and Mei{\ss}ner, Ulf-G. and Zhang, Yu and Zhang, Zhen-Hua",
    title = "{Dispersive analysis of the isospin breaking in the X(3872){\textrightarrow}J/{\ensuremath{\psi}}{\ensuremath{\pi}}+{\ensuremath{\pi}}- and X(3872){\textrightarrow}J/{\ensuremath{\psi}}{\ensuremath{\pi}}+{\ensuremath{\pi}}0{\ensuremath{\pi}}- decays}",
    eprint = "2409.13245",
    archivePrefix = "arXiv",
    primaryClass = "hep-ph",
    doi = "10.1103/PhysRevD.111.014031",
    journal = "Phys. Rev. D",
    volume = "111",
    number = "1",
    pages = "014031",
    year = "2025"
}

@article{Du:2019pij,
    author = "Du, Meng-Lin and Baru, Vadim and Guo, Feng-Kun and Hanhart, Christoph and Mei\ss{}ner, Ulf-G and Oller, Jos\'e A. and Wang, Qian",
    title = "{Interpretation of the LHCb $P_c$ States as Hadronic Molecules and Hints of a Narrow $P_c(4380)$}",
    eprint = "1910.11846",
    archivePrefix = "arXiv",
    primaryClass = "hep-ph",
    doi = "10.1103/PhysRevLett.124.072001",
    journal = "Phys. Rev. Lett.",
    volume = "124",
    number = "7",
    pages = "072001",
    year = "2020"
}

@article{Contessi:2024vae,
    author = "Contessi, Lorenzo and Pavon Valderrama, Manuel and van Kolck, Ubirajara",
    title = "{Limits on an improved action for contact effective field theory in two-body systems}",
    eprint = "2403.16596",
    archivePrefix = "arXiv",
    primaryClass = "nucl-th",
    doi = "10.1016/j.physletb.2024.138903",
    journal = "Phys. Lett. B",
    volume = "856",
    pages = "138903",
    year = "2024"
}

@article{Brambilla:2010cs,
    author = "Brambilla, N. and others",
    title = "{Heavy Quarkonium: Progress, Puzzles, and Opportunities}",
    eprint = "1010.5827",
    archivePrefix = "arXiv",
    primaryClass = "hep-ph",
    reportNumber = "SLAC-R-996, TUM-EFT-11-10, CLNS-10-2066, ANL-HEP-PR-10-44, ALBERTA-THY-11-10, CP3-10-37, FZJ-IKP-TH-2010-24, INT-PUB-10-059, JLAB-THY-11-1308, FERMILAB-PUB-10-737-T",
    doi = "10.1140/epjc/s10052-010-1534-9",
    journal = "Eur. Phys. J. C",
    volume = "71",
    pages = "1534",
    year = "2011"
}

@article{Ali:2017jda,
    author = {Ali, Ahmed and Lange, Jens S\"oren and Stone, Sheldon},
    title = "{Exotics: Heavy Pentaquarks and Tetraquarks}",
    eprint = "1706.00610",
    archivePrefix = "arXiv",
    primaryClass = "hep-ph",
    reportNumber = "DESY-17-071",
    doi = "10.1016/j.ppnp.2017.08.003",
    journal = "Prog. Part. Nucl. Phys.",
    volume = "97",
    pages = "123--198",
    year = "2017"
}

@article{Esposito:2016noz,
    author = "Esposito, A. and Pilloni, A. and Polosa, A. D.",
    title = "{Multiquark Resonances}",
    eprint = "1611.07920",
    archivePrefix = "arXiv",
    primaryClass = "hep-ph",
    reportNumber = "JLAB-THY-16-2301",
    doi = "10.1016/j.physrep.2016.11.002",
    journal = "Phys. Rept.",
    volume = "668",
    pages = "1--97",
    year = "2017"
}

@article{Guo:2017jvc,
    author = "Guo, Feng-Kun and Hanhart, Christoph and Mei\ss{}ner, Ulf-G. and Wang, Qian and Zhao, Qiang and Zou, Bing-Song",
    title = "{Hadronic molecules}",
    eprint = "1705.00141",
    archivePrefix = "arXiv",
    primaryClass = "hep-ph",
    doi = "10.1103/RevModPhys.90.015004",
    journal = "Rev. Mod. Phys.",
    volume = "90",
    number = "1",
    pages = "015004",
    year = "2018",
    note = "[Erratum: Rev.Mod.Phys. 94, 029901 (2022)]"
}

@article{Liu:2024uxn,
    author = "Liu, Ming-Zhu and Pan, Ya-Wen and Liu, Zhi-Wei and Wu, Tian-Wei and Lu, Jun-Xu and Geng, Li-Sheng",
    title = "{Three ways to decipher the nature of exotic hadrons: Multiplets, three-body hadronic molecules, and correlation functions}",
    eprint = "2404.06399",
    archivePrefix = "arXiv",
    primaryClass = "hep-ph",
    doi = "10.1016/j.physrep.2024.12.001",
    journal = "Phys. Rept.",
    volume = "1108",
    pages = "1--108",
    year = "2025"
}

@article{Godfrey:1985xj,
    author = "Godfrey, Stephen and Isgur, Nathan",
    title = "{Mesons in a Relativized Quark Model with Chromodynamics}",
    doi = "10.1103/PhysRevD.32.189",
    journal = "Phys. Rev. D",
    volume = "32",
    pages = "189--231",
    year = "1985"
}

\end{document}